\documentclass[twocolumn,floats,floatfix,aps,pra]{revtex4-2}
\usepackage{amsfonts,amssymb,amsmath}
\usepackage{color,calc,bm}
\usepackage{listings}
\usepackage{graphicx}
\usepackage{bm}
\usepackage{array}
\usepackage{hyperref}
\usepackage{physics}
\usepackage[export]{adjustbox}
\usepackage{float}
\usepackage{enumitem}
\usepackage{bm}
\usepackage{mathrsfs}
\usepackage[scr=rsfso,cal=zapfc,frak=euler,bb=ams]{mathalfa}
\usepackage{braket}  % Required for \ket command
\usepackage{xcolor}
\usepackage{quantikz}
\usepackage{tikz}
\def\be{ \begin{equation} }
\def\beal{ \begin{align} }
\def\ee{ \end{equation} }
\def\bea{ \begin{eqnarray} }
\def\eea{ \end{eqnarray} }
\def\bse{ \begin{subequations} }
\def\ese{ \end{subequations} }
\def\bwt{ \begin{widetext} }
\def\ewt{ \end{widetext} }

\def\1{\scalebox{1.5}a}
\def\2{\scalebox{1.5}b}
\def\3{\scalebox{1.5}c}

\def\C{C}

\def\A{\mathbf{A}}
\def\B{\mathbf{B}}
\def\C{\mathbf{C}}
\def\E{\pmb{\mathcal{E}}}
\def\ea{\pmb{\mathcal{E}^2_{a}}}
\def\eb{\pmb{\mathcal{E}^2_{b}}}
\def\en{\pmb{E}^{(N)}}
\def\R{\pmb{\mathcal{R}}}
\def\U{\pmb{\mathcal{U}}}
\def\rm{\pmb{\mathcal{R}}_{N}}

\def\T{\pmb{\mathcal{T}}}
\def\I{\pmb{\mathcal{I}}}

\def\M{\pmb{\mathcal{\R}}^N}

\def\L{\pmb{\mathscr{S}}}
\def\hence{\Longrightarrow}

\def\MPPT{MQPT}

\begin{document}

\title{Multipass Quantum Process Tomography: Precision and Accuracy Enhancement}

\author{Stancho G. Stanchev and Nikolay V. Vitanov}

\affiliation{Center for Quantum Technologies, Faculty of Physics, St Kliment Ohridski University of Sofia, 5 James Bourchier blvd, 1164 Sofia, Bulgaria}

\date{\today }

\begin{abstract}
We introduce a method to enhance the precision and accuracy of Quantum Process Tomography (QPT) by mitigating the errors caused by state preparation and measurement (SPAM), readout and shot noise. 
Instead of performing QPT solely on a single gate, we propose performing QPT on a sequence of multiple applications of the same gate. 
%By using the Pauli transfer matrix of the multipass process, we propose a post-processing procedure for a more precise and accurate characterization of the single process, which greatly reduces the SPAM, readout and shot noise errors. 
The method involves the measurement of the Pauli transfer matrix (PTM) by standard QPT of the multipass process, and then deduce the single-process PTM by two alternative approaches: an iterative approach which in theory delivers the exact result for small errors, and a linearized approach based on solving the Sylvester equation.
We examine the efficiency of these two approaches through simulations on IBM Quantum using \textsc{ibmq\_qasm\_simulator}.
Compared to the Randomized Benchmarking type of methods, the proposed method delivers the entire PTM rather than a single number (fidelity). 
Compared to standard QPT, our method delivers PTM with much higher accuracy and precision because it greatly reduces  the SPAM, readout and shot noise errors.
We use the proposed method to experimentally determine the PTM and the fidelity of the CNOT gate on the quantum processor \textsc{ibmq\_manila} (Falcon r5.11L). %~\cite{IBM}.
\end{abstract}

\maketitle

%%%%%%%%%%%%%% SEC %%%%%%%%%%%%%% SEC %%%%%%%%%%%%%% SEC %%%%%%%%%%%%%% SEC %%%%%%%%%%%%%% 
%%%%%%%%%%%%%% SEC %%%%%%%%%%%%%% SEC %%%%%%%%%%%%%% SEC %%%%%%%%%%%%%% SEC %%%%%%%%%%%%%% 
%%%%%%%%%%%%%% SEC %%%%%%%%%%%%%% SEC %%%%%%%%%%%%%% SEC %%%%%%%%%%%%%% SEC %%%%%%%%%%%%%% 
\section{Introduction}\label{Sec:Intro}

Quantum Process Tomography (QPT) is a crucial technique in quantum information processing, which provides valuable insights into the behavior of quantum systems and their constituent quantum operations~\cite{NielsenChuang2012, ChuangNielsen1997,Poyatos1997, O'Brien2004}. 
In the pursuit of large-scale quantum computers, it plays a vital role in characterizing the fidelity of quantum gates and quantum circuits~\cite{Riebe2006, Tinkey2021, Bialczak2010}.
However, QPT and in particular Standard QPT (SQPT), face significant challenges. 
They typically require a large number of measurements in order to accurately characterize a quantum process. 
As the size of the quantum system grows, the number of required measurements increases exponentially, making the process resource-intensive and time-consuming. 
Moreover, in order to achieve high precision and accuracy, a huge number of measurement shots is required, with minimal SPAM and readout errors.

Recent advancements in both theory and practical applications have significantly improved the quality of quantum process tomography (QPT). 
From the perspective of how the characterized process is applied within the circuit, prior to the measurement, we can categorize these techniques into two groups: those that employ single process applications and those that utilize multiple applications, often referred to as long-gate sequence, gate repetition, concatenation,  multipass, etc.  

The single-process application techniques include, for example, Standard QPT~\cite{NielsenChuang2012, ChuangNielsen1997,Poyatos1997, O'Brien2004,Fiurasek2001}, Ancilla-Assisted QPT (AAPT)~\cite{Altepeter2003}, Direct Characterization of Quantum Dynamics (DCQD)~\cite{Mohseni2008, Roncallo2024}, Compressed sensing QPT~\cite{Rodionov2014},  Gradient-Descent QPT~\cite{Ahmed2023}, Self-Consistent QPT~\cite{Merkel2013, Sugiyama2021}, Projected Least-Squares QPT~\cite{Surawy-Stepney2022}, etc. These methods, especially SQPT, are known to be relatively susceptibility to SPAM and readout errors.  
The methods that greatly improve the precision of tomography are those that use gate repetitions. 
The repetitions amplify the structured and randomized gate errors and reduce the effects of systematic and/or statistical measurement errors caused by SPAM, readout and shot noise. 
The most widely adopted such protocols are Randomized Benchmarking (RB), Interleaved RB, and Gate Set Tomography (GST). 
Some other examples are discussed in Refs.~\cite{Chow2012a, Kunjummen2023, Gulliksen2015}.

Randomized benchmarking~\cite{Knill2008, Magesan2011, Proctor2017, Magesan2012, Emerson2005, Erhard2019, Corcoles2013, Cross2019} entails executing prolonged sequences of random Clifford group gates that combine to create the identity, enabling the observation of performance degradation as the circuit depth increases. 
However, RB serves as a protocol for characterizing the average performance. 
It demonstrates efficient scalability with respect to the number of qubits affected by the characterized gate set and remains resilient in the presence of noise during state preparation and measurement.
Interleaved RB~\cite{Magesan2012a} is designed to characterize a particular gate, which is ‘interleaved’ throughout the standard RB sequence. The protocol is a simple extension of standard RB which involves running sequences of random gates.
In GST~\cite{Nielsen2021, Greenbaum2015}, the key distinction in comparison to older QPT methods is its calibration-free nature. 
Unlike its predecessors, GST does not rely on pre-calibrated state preparations and measurements. However, obtaining a GST estimate entails solving a highly nonlinear optimization problem. Moreover, the scaling with system size is polynomially worse than QPT due to the necessity to characterize multiple gates simultaneously. In order to estimate a complete gate set, approximately 80 experiments are required for single-qubit gates and over 4000 for two-qubit gates. 
This is far worse than QPT, which only requires 12 experiments for a single-qubit gate and 144 experiments for a two-qubit gate~\cite{Greenbaum2015}.
{There are also methods, as the one discussed in ~\cite{Chow2012b} that utilize gate application of a sequence of identical repetitions}.

\begin{figure}[tb]
    \makebox[\linewidth]{
\includegraphics[width=0.85\columnwidth]{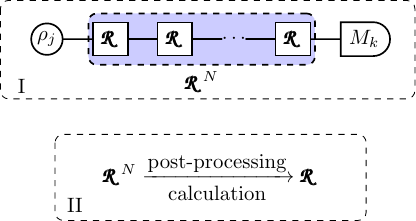}
    }

\caption{Multi-pass Process Tomography (\MPPT) comprises two distinct stages. Stage I involves a conventional Process Tomography (e.g., SQPT, AAPT, etc.), but instead of measuring a single process $\R$, it characterizes a multi-pass process $\R^N$, which consists of $N$ identical processes (gates). In Stage II, the measured multi-pass process is used to extract the single process through a post-processing calculation procedure.}
\label{Fig:RepProcess}
\end{figure}

In this paper, we introduce a method called \emph{Multipass Quantum Process Tomography} (\MPPT~) that combines the methods based on repetitions with those that use single-gate application only.  
Contrary to most repetition methods, our method uses only repetitions of \textit{the same gate}.  
The repetitions lead to the amplification of the single-pass errors in the \emph{multipass} errors, which are then measured with high accuracy and precision by standard QPT.
There are two clear advantages of using only the same gate (see Fig.~\ref{Fig:RepProcess}): (i) we do not introduce additional errors stemming from other gates, and (ii) it is possible to design a simple and very fast post-processing procedure that derives the individual process from the multipass process.    
Our method is distinguished by its intuitiveness and simplicity, as it uses the already developed single-pass techniques, such as SQPT, and reduced mathematical complexity, making it both theoretically accessible and practically feasible. 
Yet, our method features significant accuracy and precision enhancement compared to single-pass SQPT and potentially to other single-pass methods.

%%%%%%%%%%%% Fig %%%%%%%%%%%% Fig %%%%%%%%%%%% Fig %%%%%%%%%%%% Fig %%%%%%%%%%%%

%%%%%%%%%%%% Fig %%%%%%%%%%%% Fig %%%%%%%%%%%% Fig %%%%%%%%%%%% Fig %%%%%%%%%%%%

The paper is organized as follows. 
Section \ref{MQPT} provides an overview of single and multi-pass processes, followed by a description of the conceptual framework of~\MPPT. 
In Sec.~\ref{Iter_Lin}, our Iterative and Linear methods of \MPPT~are presented. 
Section \ref{Dem_sim} demonstrates the application of \MPPT~methods to $\sqrt{X}$ and CNOT gates, respectively, through simulations on the IBM Quantum simulator.  
In Sec.~\ref{ExpDemManila}, we demonstrate the methods on a real IBM Quantum processor. 
Finally, in Sec.~\ref{Conclusions}, we present our conclusions.

%%%%%%%%%%%%%% SEC %%%%%%%%%%%%%% SEC %%%%%%%%%%%%%% SEC %%%%%%%%%%%%%% SEC %%%%%%%%%%%%%% 
%%%%%%%%%%%%%% SEC %%%%%%%%%%%%%% SEC %%%%%%%%%%%%%% SEC %%%%%%%%%%%%%% SEC %%%%%%%%%%%%%% 
%%%%%%%%%%%%%% SEC %%%%%%%%%%%%%% SEC %%%%%%%%%%%%%% SEC %%%%%%%%%%%%%% SEC %%%%%%%%%%%%%% 
\section{Multipass Quantum Process Tomography (\MPPT)}\label{MQPT}

\subsection{Motivation}

As noted in Introduction, a sequence of repetitions is widely used in the process characterization due to the fact that the errors are amplified (accumulated) as the number of repetitions $N$ increases. 
These larger multipass errors are more easily measurable with high precision and accuracy than the small single-pass errors.  
The single-pass errors can be deduced from the multipass signal provided that we have an unambiguous connection between the single-pass and multipass processes. 
The latter requirement imposes some limitations on the formalism which can be used to describe the multipass quantum process.

In the unitary case, due to the absence of decoherence, the error amplification can even be quadratic \cite{Vitanov2020}. 
Then it is relatively easy to find the $N$th power of a given parameterized propagator, as has been done in several works~\cite{Vitanov1995,Stanchev2023, Stanchev2024, Vitanov2018, Baldwin2014, Zhou2015}. 
It is also possible to find the single propagator from the $N$-pass propagator (its $N$th power), which makes it possible to accurately and precisely deduce the single-pass propagator errors from the amplified multi-pass errors. 
If the propagator $U$ is explicitly defined with respect to time during a time period from 0 to the gate time $T$, then exponentiating $U^N$ can be easily obtained by scaling $T$ to $NT$, for example, in the M{\o}lmer-S{\o}rensen gate~\cite{Sorensen2000}.  

However, if the evolution is not unitary and it is instead represented by a Completely Positive and Trace-Preserving (CPTP) map, as it occurs when considering open quantum system noise, then finding the $N$th power of a process by parameterizing the single process becomes a challenging task. 
For the sake of process tomography, we are very interested in the inverse task, i.e., finding the first power from the known $N$th power, which is also a daunting task. 
The extraction of the single CPTP process from the amplified multi-pass process will be referred to as \emph{Multipass Quantum Process Tomography} (\MPPT). 

In this paper, we will treat the terms ``propagator'', ``gate'', ``process'', and ``channel'' as synonymous concepts, recognizing that they are often used interchangeably or may refer to similar aspects of quantum information and quantum computing.

\subsection{Single process and distance metrics}

\subsubsection{Channel characteristics}

Before delving into multipass processes, let us first describe the overall concept of a process. 
A quantum process (channel) is a quantum operation represented by a CPTP map. 
Various CPTP map representations ensure the physicality of quantum processes by preserving properties such as trace and positivity \cite{Choi1972, Choi1975, Jamiolkowski1972} and adhere to Markovian dynamics \cite{Lindblad1976, Gorini1976}. 
Examples of such representations are the Choi matrix,  $\chi$-Process matrix, Kraus (Operator sum representation)~\cite{Kraus1983}, Stinespring~\cite{Stinespring1955}, Liouville (Superoperator), Pauli Transfer matrix (PTM)~\cite{Merkel2013, Gambetta2013}, etc. 
All of them are equivalent and can be converted into each other~\cite{Wood2015}. 
In our analysis, we use  PTM ($\R$) and Liouville ($\L$), because if a given process is repeated $N$ times then the resultant multipass process is simply  $\R^N$ and $\L^N$, respectively.
These two representations govern the evolution of the vectorized density matrix $\ket{\rho}\!\rangle$. 
The density matrix vectorization in Liouville representation is simply the column stacking of $\rho$, i.e. 
\be
|\rho\rangle\rangle:=\left(\rho_{11}, \rho_{21}, \ldots, \rho_{k 1}, \rho_{12}, \rho_{22}, \ldots, \rho_{k 2}, \rho_{kk }\right)^T,
\ee
while the corresponding  channel $\L$ is a complex matrix. 
On the other hand, the PTM $\R$ is a real matrix with elements in the range from $-1$ to $1$ and the respective density matrix vectorization is in the $n$-qubit Pauli basis, denoted by $\mathcal{P}^{\otimes n}$. 
It forms an operator basis for a $d=2^n$ dimensional Hilbert space, encompassing a total of $d^2=4^n$ operators. 
These operators can be conveniently labeled using a single index, such as $P_1=I^{\otimes n}$, $P_2$, $P_3$, up to $P_{d^2}=Z^{\otimes n}$. These operators are Hermitian and unitary, satisfying the property $P_k^2=I^{\otimes n}$.
In the case of a single qubit ($n=1$), the basis is represented by $\mathcal{P}^{\otimes 1}=\{{I}, {X}, {Y}, {Z}\}$, which correspond to the familiar Pauli matrices.
Expanding to two qubits ($d=4$), we form a set of 16 operators:  $\{I I, I X, I Y, I Z, X I, X X, X Y, \ldots, Z Z\}$, where $I I$ should be interpreted as $I \otimes I$, and so on. 
These operators are succinctly denoted as $\left\{P_1, P_2, \ldots, P_{16}\right\}$ ~\cite{Forest2019}.

The density matrix vectorization for the single qubit reads:
\be
\rho=\frac{1}{2}\left(\begin{array}{cc}1+z & x-i y \\ x+i y & 1-z\end{array}\right) \quad \hence \quad |\rho\rangle\rangle=\left(\begin{array}{l}1 \\ x \\ y \\ z\end{array}\right),
\ee
%\textcolor{blue}
{where $x,y,z$ are the Bloch vector components}.
%As in the Liouville case,
In the PTM representation, the evolution of the state can be expressed as 
\be
\ket{\rho}\!\rangle = \R\ket{\rho_0}\!\rangle,
\ee
where $\ket{\rho_0}\!\rangle$ is the initial vectorized state.

The conversion from PTM to Liouville representation is done by the transformation
\be
\L=\U^{\dagger}\R \U,
\ee 
where $\U$ is the unitary operator 
\be
\U=\sum_{k=1}^{d^2}|c_k\rangle\langle\langle P_k|,
\ee
where all $|c_k\rangle$ represent the computational basis and they are all possible strings from $|c_1\rangle = |0,0,\ldots,0\rangle$ to $|c_{d^2}\rangle = |1,1,\ldots,1\rangle$.

For the purposes of quantum computation, the channels must have unitary targets $\T$ (or $\T_{L}$), represented in PTM and Liouville representations, respectively. 
These targets can be expressed by the corresponding $d\times d$ target unitary operator $\bm{T}$ as
\bse
\begin{align}
\T &=\U \Big(\bm{T}^* \otimes \bm{T} \Big)\U^{\dagger}, \\
\T_{L} &=\bm{T}^* \otimes \bm{T}.
\end{align}
\ese 
For the ensuing analysis, we will employ the PTM representation, as it is equivalent to the Liouville representation but it is more convenient for our objectives.

We define the discrepancy between the actual process $\R$ and the target process $\T$ as the error matrix
\be \label{errorM1} 
\E=\R-\T,
\ee
where $\E$ in general is not a CPTP matrix. 
The characterization and analysis of $\E$ play a central role in evaluating the accuracy and precision of the process tomography methods discussed hereafter.

\subsubsection{Distance metrics}

Given the matrices $\R$, $\T$ and $\E$, we can calculate the process distance metrics. 
In this paper, we utilize the \textit{diamond norm} alongside process infidelity metrics. 
The diamond norm (known also as \textit{completely bounded trace norm})~\cite{Aharonov1998, Kitaev1997} is one of the most effective measures to assess the proximity of two processes due to its enhanced rigor. 
In our analysis, we formally express it as  
\be
\|\E\|_{\diamond}=\|\R-\T\|_{\diamond}.
\ee

For more comprehensive details about the diamond norm, its definition and calculation, we refer the reader to Refs.~\cite{Watrous2009, Watrous2013}.
In our work, for the numerical calculation of $\|\E\|_{\diamond}$ and the CPTP channel representations and transformations, we utilize the following Python ready-to-use tools:
\textsc{qiskit.quantum\_info},~\cite{Qiskit},
\textsc{forest.benchmarking.operator\_tools}~\cite{Forest2019}, and
\textsc{qutip.superop\_reps}~\cite{Qutip2013}.

The process fidelity $F$ and infidelity $e_F$, %\textcolor{blue}
{(when $\T$ is unitary ~\cite{Chow2012b})} are 
\bse\label{infidelity1}
\begin{align}
F& = \frac{\operatorname{Tr}[\T^{\dagger}\R]}{d^2}= 1 + \frac{\operatorname{Tr}[\T^{\dagger}\E]}{d^2}, \\
e_F&=1-F =- \frac{\operatorname{Tr}[\T^{\dagger}\E]}{d^2}.
\end{align}
\ese
%where $\I$ is the identity matrix.

Measurement of $e_F$ is achievable through Interleaved RB.
However, there is currently no established scalable method for determining $\|\E\|_{\diamond}$. 
Tomographic methods, such as GST, can be utilized to estimate $\|\E\|_{\diamond}$, but they become exponentially expensive as the number of qubits $n$ increases. 
As discussed in Ref.~\cite{Hashim2023}, if only $e_F$ is known, it is possible to bound $\|\E\|_{\diamond}$ using 
\be \label{bound}
e_F \leq \|\E\|_{\diamond} \leq d\sqrt{e_F}, \ee which means that the larger number of qubits $n$ $(d = 2^n)$ leads to larger inaccuracy of $\|\E\|_{\diamond}$. 
The objective of the \MPPT~method is not only to measure $\|\E\|_{\diamond}$ more accurately, but also to find the individual elements of the error matrix $\E$. 
Therefore, for our simulations, we introduce an additional metric called \textit{differential diamond norm} that is determined by the difference between the measured error matrix $\E_N$ and the actual one, i.e. $\|\E_N-\E\|_{\diamond}$.

\subsection{\MPPT: conceptual framework}

%%%%%%%%%%%% Fig %%%%%%%%%%%% Fig %%%%%%%%%%%% Fig %%%%%%%%%%%% Fig %%%%%%%%%%%%

The basic structure of \MPPT~is shown schematically in Fig.~\ref{Fig:RepProcess} and it assumes in general two stages. 
In stage I, $N$ processes (passes) are applied instead of a single process $\R$, so that the sequence $\R^N$, which represents the multipass process, is the object of characterization. 
The aim is to amplify the gate error $\E$ and use the enhanced precision and accuracy of the measured multipass process in order to enhance the precision and accuracy of the deduced single-pass process. 
The multipass process can be characterized by a conventional method, such as SQPT, AAPT, or any other methods that in principle apply the characterizing process only once. 
Because the multipass process features large (accumulated) errors, the latter can be determined with high accuracy and precision.
Subsequently, stage II is a post-processing calculation procedure used to obtain the single process $\R$ from the multipass process $\R^N$. 
Stage I includes measurement errors (SPAM, readout, shot noise), introduced by the preparation and projective measurement operations $\rho_j$ and $M_k$. 
All measurement errors remain constant at both ends of the entire circuit, i.e. they do not accumulate as $N$ increases and they are presumed to be the same for single-pass and multipass processes. 
This leads to an improvement in the measurement of the multipass process $\R^N$, since only the gate error $\E$ accumulates over its course.

%\textcolor{blue}
Before proceeding to stage II, let us make a rough estimate of the resource increase with such an approach. In contemporary quantum superconducting processors, the time for a single shot $t_s$ (even in the absence of any gate in the circuit) is about 0.3 ms, whereas the time for a single pass (gate) $t_g$ is on the order of $\sim 100$ ns and $\sim600$ ns for single-qubit and two-qubit gates, respectively. 
This means that with a number of passes $N \approx 20$, the time for a single shot (for the two-qubit case) will be $t'_s =t_s + Nt_g\approx1.04t_s$, so that it will increases by only about 4\% compared to the application of a single pass. 
This implies that the method does not significantly increase the cost of the experiment in practice.

For performing stage II --- deducing the single-pass PTM from the multipass PTM --- we use two approaches: 

\begin{itemize}
\item an iterative method;

\item a linear approximation.
\end{itemize}
The iterative method delivers a solution that is practically exact regarding the input data derived from the measured matrix $\M$. 
In this context, the term 'exact' refers solely to the correspondence between the input and the output, rather than indicating precision in identifying the actual error matrix $\E$.  
However, if $\M$ is considered to be measured with small errors, i.e., with a large number of shots, then initial measurement errors (such as SPAM and readout) are mitigated without introducing any additional post-processing errors. 
This leads to a significant enhancement in both precision and accuracy in the deduced single-pass process.

The linear method introduces some post-processing errors due to considering only the linear terms in the expansion of $\M$ with respect to the error matrix $\E$. 
Hence, it applies to small errors only. 
Despite this, the method also proves to be useful, especially when the tomography is performed with relatively small number of shots, i.e. in the presence of large shot noise. 
In this scenario, it positively affects precision and accuracy, showing some benefits over the iterative method. 
This can be likened to the principles of a Wiener filter in signal and image processing~\cite{Boulfelfel1994}, where a linear approach is favored for its ability to effectively reduce noise and enhance the clarity of imaging in noisy environments.

%%%%%%%%%%%%%% SEC %%%%%%%%%%%%%% SEC %%%%%%%%%%%%%% SEC %%%%%%%%%%%%%% SEC %%%%%%%%%%%%%% 
%%%%%%%%%%%%%% SEC %%%%%%%%%%%%%% SEC %%%%%%%%%%%%%% SEC %%%%%%%%%%%%%% SEC %%%%%%%%%%%%%% 
%%%%%%%%%%%%%% SEC %%%%%%%%%%%%%% SEC %%%%%%%%%%%%%% SEC %%%%%%%%%%%%%% SEC %%%%%%%%%%%%%% 
\section{Iterative and Linear methods}\label{Iter_Lin}

%This section represents the theoretical results.

As discussed above, the second stage of \MPPT~is the inverse task, in which the objective is to extract  the single process $\R$, and respectively, the error matrix $\E$, from the multipass process $\M = (\T+\E)^N$. 
Generally speaking, this problem has many solutions.
However, assuming that $\R$ is a high-fidelity gate, i.e., $\|\E\| \ll \|\R\|$, and $\T$ is a known target matrix, then we can find a definitive single solution. 
We will do that first considering the practically exact iterative solution and then we present the linear method.

\subsection{Post-processing iterative method}

The method is based on an iterative %perturbative 
approach. 
%This technique focuses on refining a matrix $\E$ to align the $N$-th power of the sum of $\E$ and a target matrix $\T$ with a pre-measured matrix $\M$. 
It combines the principles of perturbative methods with iterative refinement, akin to gradient descent methods.
At its core, the algorithm employs a perturbative strategy, introducing $\E$ as a minor modification to $\T$. This perturbation, significantly smaller in magnitude compared to $\T$ ($\|\E\| \ll \|\T\|$), ensures the preservation of the fundamental characteristics of $\T$, providing stability to the solution process.
The method unfolds in an iterative cycle, where each iteration corrects $\E$ based on the difference between $(\T + \E)^N$ and $\M$. This iterative refinement directs the solution towards a high degree of precision, typically reducing the error to less than $10^{-12}$, achieving what is called a 'practically exact' solution in the field of numerical computations.

%\subsection*{Iterative Method Description}

The iterative method proceeds as follows: 
\begin{itemize}
\item \emph{Initialization:} Start with an initial guess for $\E$, usually $\E = 0$.

\item \emph{Iterative Update Steps:}

(i) Calculate $\M_{\text{current}} = (\T + \E)^N$.

(ii) Determine the difference: $\bm{\delta} = \M - \M_{\text{current}}$.

(iii) Update $\E$ using: $\E = \E + \alpha  \bm{\delta}$, where $\alpha$ is a small positive learning coefficient (e.g., 0.01).

\item \emph{Convergence Criterion:} The process repeats until the norm of $(\T + \E)^N - \M$ falls below a specified tolerance (e.g., $10^{-12}$), indicating that a practically exact solution has been found.

\end{itemize}
The code is presented in appendix~\ref{python code}.

It is important to note that for the CNOT gate, the method works with $N=2m+1$ passes, while for the $\sqrt{X}$ gate it works for $N=4m+1$ passes, where $m=0,1,2,\ldots$
In section~\ref{Dem_sim}, we demonstrate the performance of this method for the $\sqrt{X}$ and CNOT gates by simulation on IBM Quantum \textsc{ibmq\_simulator}.

\subsection{Linear method for involutary gates}

Now we will derive the linear approximation for the solution of $\E=f(\M,\T,N)$, which is valid for involutary gates, i.e., gates which satisfy the relation
\be\label{involutory}
\T^2=\I,%\bm{\mathbb{I}}.
\ee
Some examples for involutory gates are
\begin{itemize}
\item single qubit gates: $X$, $Y$, $Z$ and Hadamard;

\item two-qubit gates: CNOT, SWAP, Controlled Z;

\item three-qubit gates: Toffoli and Fredkin.
\end{itemize}

For these gates, it is convenient to choose the number of the passes to be $N=2m+1$ $(m=0,1,2,\ldots)$, so we have $\T^{2m+1}=\T$. % and the integer number $m$ will be called \emph{cycles}.   
Then the multipass process 
\be
\R^{2 m+1}= (\T+\E)^{2m+1}
\ee
reads, when expanded up to the second power of $\E$, as 
\begin{subequations}\label{UnE2}
\begin{align}
\R^{2 m+1} &= \T+(m+1) \E+m \T\E \T \label{UnE2-1} \\
&+\frac{m(m+1)}{2}\ea + \frac{m(m-1)}{2}\eb + O(\E^3),
\end{align}
\end{subequations}
with
\bse\label{Eab}
\begin{align}
\ea&=\T\E^2+\E\T \E + \E^2\T , \\
\eb&=\T\E\T \E\T,
\end{align}
\ese 
where we have used the non-commutative rules and we have accounted for Eq.~\eqref{involutory}.

%If the error matrix $\E$ is known, Eq.~\eqref{UnE2} exhibits very high accuracy with respect to the result $\R^{2 m+1}$. 
%However, we face a different challenge when $\R^{2 m+1}$ is known but $\E$ is unknown. 
In order to find a unique linear solution, we retain linear terms, i.e. Eq.~\eqref{UnE2-1}. 
The nonlinear terms $\ea$ and $\eb$ can be viewed as post-processing errors; their significance grows quadratically with $m$.
This fact sets an upper limit of how many passes we can apply before our solution deviates too much from the actual one.

%The linearized version of Eq.~\eqref{UnE2} becomes
%\be\label{linear eq}
%\R^{2 m+1} = \T+(m+1) \E + m \T\E \T.
%\ee
By multiplying Eq.~\eqref{UnE2-1} with $\T$ from the left and taking into account Eq.~\eqref{involutory}, we obtain
\be\label{Sylvester0}
\T\R ^{2 m+1}-\I=(m+1)\T \E+m\E \T.
\ee
This is the Sylvester equation,
\be\label{Sylvester1}
 \A\E + \E\B = \C,
\ee
with
\bse
\begin{alignat}{2}
\A &=(m+1)\T, \\
\B &=m \T , \\
\C &=\T\R ^{2 m+1}-\I ,
\end{alignat}
\ese
which are known matrices. 

Equation~\eqref{Sylvester1}  has a unique solution for $\E$ if and only if the matrices $\A$ and $-\B$ do not share any eigenvalue. 
In our case, this condition is always fulfilled because $m+1 \neq -m$.
This means that by solving Eq.~\eqref{Sylvester1} we can calculate the error matrix  $\E$ %=f(\M,\T,N)$
and then the single process $\R =\T+\E$.

\subsection{Linear approximation for arbitrary target gates}\label{Gen_case}

As for the involutary target gates, we proceed with the same linear approximation for the generalization for any arbitrary target gate. 
By keeping only the linear terms in $\E$, the multipass process $\M$ takes the form 
\be \label{UN_gen}
\M \approx \T^N+\T^{N-1} \sum_{s=0}^{N-1}\T^{-s}\E \T^s. 
\ee
%By putting $\T^N$ in the left hand side and multiplying with $\T^{1-N}$ from the left, Eq.~\eqref{UN_gen} can be converted to
We find from here
\be \label{EN_gen}
\en = \sum_{s=0}^{N-1}\T^{-s}\E \T^s, 
\ee
where $\en =\T^{1-N}\M - \T $ is  a known matrix %since $\M$ is known from the first stage of \MPPT,
and our aim is to find the error matrix $\E$ on the right-hand side. 
Equation~\eqref{EN_gen} is a special case of a general extended Sylvester equation.
According to Ref.~\cite{Song2011} it can be numerically solved for $\E$ by the iterations
\be \label{Ek}
\E_{k+1}=\E_{k}+\mu \sum_{q=0}^{N-1}\T^{q}\Big(\en-\sum_{s=0}^{N-1}\T^{-s}\E_{k}\T^s \Big) \T^{-q},
\ee
where $\mu$ is a convergence factor that controls the convergence of the iterative  algorithm~\eqref{Ek}, which  takes values in the range of $0<\mu\ll 1$.
Larger $\mu$ leads to fewer iterations; however, above a critical value there is no convergence of the algorithm. 
More details about the choice of $\mu$ can be found in Ref.~\cite{Song2011}. 
In our demonstration for single-qubit non-involutary gates, we choose $\mu=0.003$. 
The initial matrix $\E_0$ could be any arbitrary matrix, having the same dimension as $\T$. 

In Sec.~\ref{Dem_sim}, we demonstrate the linear method for the $\sqrt{X}$ and CNOT gates by simulations on \textsc{qasm\_simulator}.   
We will illustrate also how the linear  method leads to improved accuracy  for small number of shots $n_s$ compared to the iterative method. 
However, for large $n_s$, the post-processing errors become significant and then the iterative method outperforms the linear method.

%%%%%%%%%%%%%% SEC %%%%%%%%%%%%%% SEC %%%%%%%%%%%%%% SEC %%%%%%%%%%%%%% SEC %%%%%%%%%%%%%% 
%%%%%%%%%%%%%% SEC %%%%%%%%%%%%%% SEC %%%%%%%%%%%%%% SEC %%%%%%%%%%%%%% SEC %%%%%%%%%%%%%% 
%%%%%%%%%%%%%% SEC %%%%%%%%%%%%%% SEC %%%%%%%%%%%%%% SEC %%%%%%%%%%%%%% SEC %%%%%%%%%%%%%% 
\section{Performance of the \MPPT~method: simulation}\label{Dem_sim}
The simulations discussed in this section were performed with \texttt{qasm\_simulator} backend of Qiskit, IBM's open-source quantum computing software framework ~\cite{Qiskit}.
\subsection{Description of the simulation}

In order to examine the precision and the accuracy of the proposed method and its capability to identify individual error matrix elements, we have chosen a purely simulation-based approach. 
This choice is due to the fact that in simulations, we are able to know a priori the actual process we wish to characterize and use it as a reference. 
In order to obtain reliable statistics, we apply the iterative method to the diamond norm by performing 50 tomographies for various passes $N$ and numbers of shots $n_s$. 
We perform such simulations on two basic quantum gates --- the single-qubit gate $\sqrt{X}$ and the two-qubit gate CNOT. 
For both of them, we have pre-generated PTM matrices $\R$ that have similar gate infidelities as the respective real quantum processors of IBM Quantum. 
{The PTM's were generated using the Qutip's master equation \textsc{mesolve} program ~\cite{Qutip2013} with the incorporation of collapse operators and unitary errors that were randomly selected.}  

For the simulation of measurement, it is necessary also to simulate the measurement errors --- specifically, SPAM, readout, and shot noise errors. 
As for the gate error matrix, the measurement errors are chosen by considering typical values for the current real IBM Quantum processors. 
In our model, the SPAM error is introduced by the so-called "noise model"~\cite{Noise2024}. 
The SPAM error is modeled only for the $\sqrt{X}$ gate, as it is the only imperfect single qubit gate that is used in SPAM. 
The phase gates, which are also basic gates and used in SPAM, are practically perfect due to the usage of the so-called "Frame change" technique~\cite{McKay2017}. 
We assume a typical SPAM error (infidelity) of $2\times 10^{-4}$ for the $\sqrt{X}$ gate, while the readout error is set to $3\times10^{-3}$. 
We note that, in reality, the readout error could be larger, in the range of $10^{-2}$; however, we assume that we have also performed in advance the so-called "readout mitigation" technique~\cite{MPT2024}. 
This technique is widely used and is available in Qiskit in the module \textsc{MitigatedProcessTomography}. Nevertheless, this technique cannot completely mitigate the readout error, and a smaller but still significant part remains.  
The shot noise error, that scales proportionally to $1/\sqrt{n_{s}}$ is automatically introduced when specifying the number of shots $n_{s}$ in the Qiskit \textsc{ProcessTomography} module. 
In the simulation discussed in this section, we use various numbers of shots $n_s=$4k, 10k, 40k, 100k etc., where, for simplicity, k$=10^3$ hereafter.

%%%%%%%%%%%% Fig %%%%%%%%%%%% Fig %%%%%%%%%%%% Fig %%%%%%%%%%%% Fig %%%%%%%%%%%%
\begin{figure}[tbph]
    \makebox[\linewidth]{
\includegraphics[width=0.9\columnwidth]{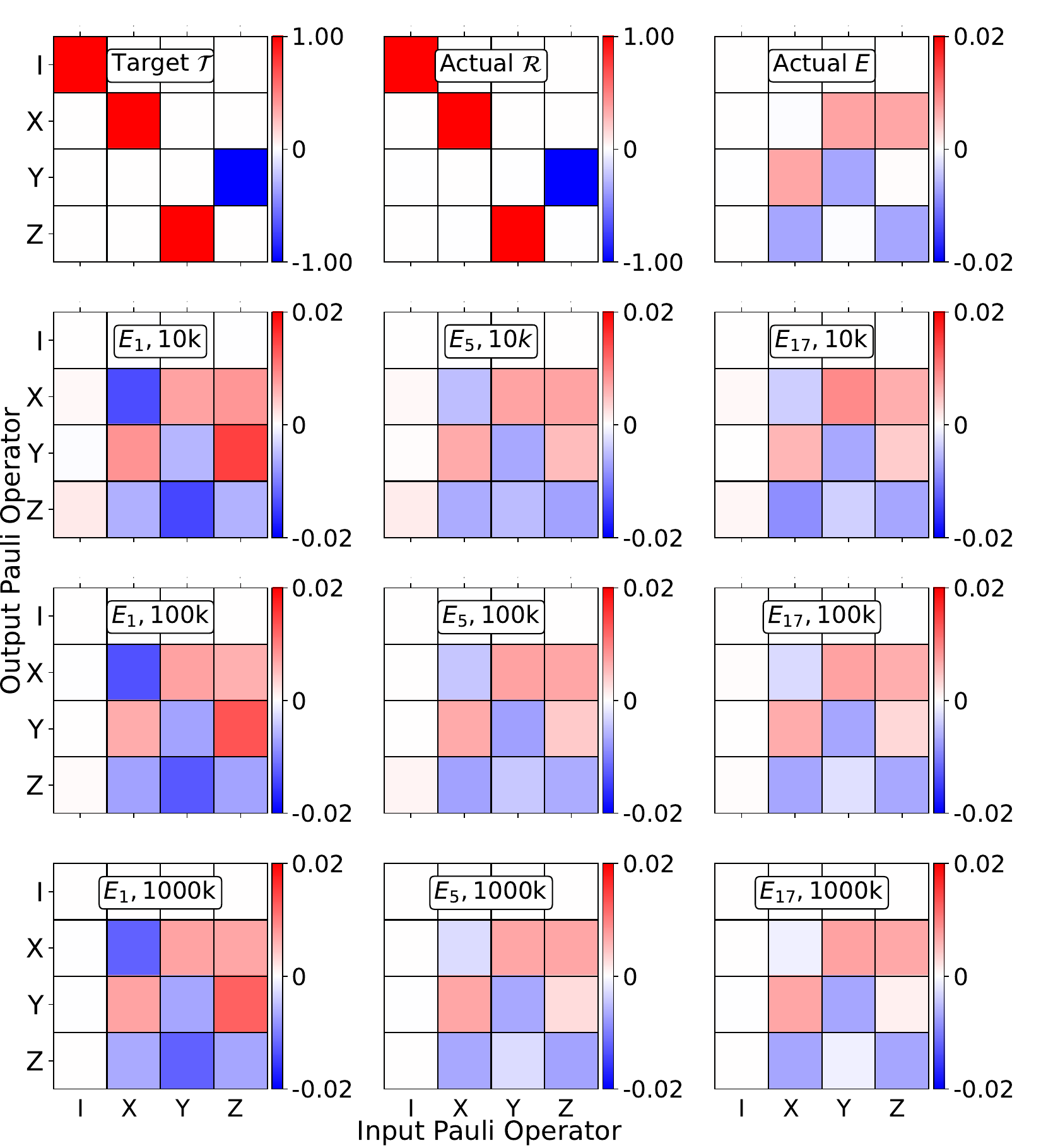}
    }
\caption{
\textit{Top:} Target PTM $\T$  for the $\sqrt{X}$ gate (left), actual (simulated) PTM $\R$ (middle), actual (simulated) error matrix $\E$ (right) with a diamond norm of $\| \E \|_{\diamond} = 0.018$, infidelity $e_F=0.00019$. 
%The actual process $\R$ is the sum of these both matrices.
\textit{Bottom 3 rows:} Tomography of the $\sqrt{X}$ gate by the iterative \MPPT~method,  represented by the measured error matrices $\E_N$ corresponding to various numbers of passes $N=1,5,17$ and various numbers of shots (10k, 100k and 1000k). 
Standard single-pass QPT corresponds to $\E_1$ ($N=1$, left column).  
These error matrices are averaged over 50 individual tomographies. 
}
     \label{Fig:SX_PTM_E_En1}
\end{figure}
%%%%%%%%%%%% Fig %%%%%%%%%%%% Fig %%%%%%%%%%%% Fig %%%%%%%%%%%% Fig %%%%%%%%%%%%

%This will provide us with an initial understanding of the method's reliability.

%%%%%%%%%%%% Fig %%%%%%%%%%%% Fig %%%%%%%%%%%% Fig %%%%%%%%%%%% Fig %%%%%%%%%%%%
\begin{figure*}[tbph]
\begin{tabular}{cc}
\includegraphics[width=0.9\columnwidth]{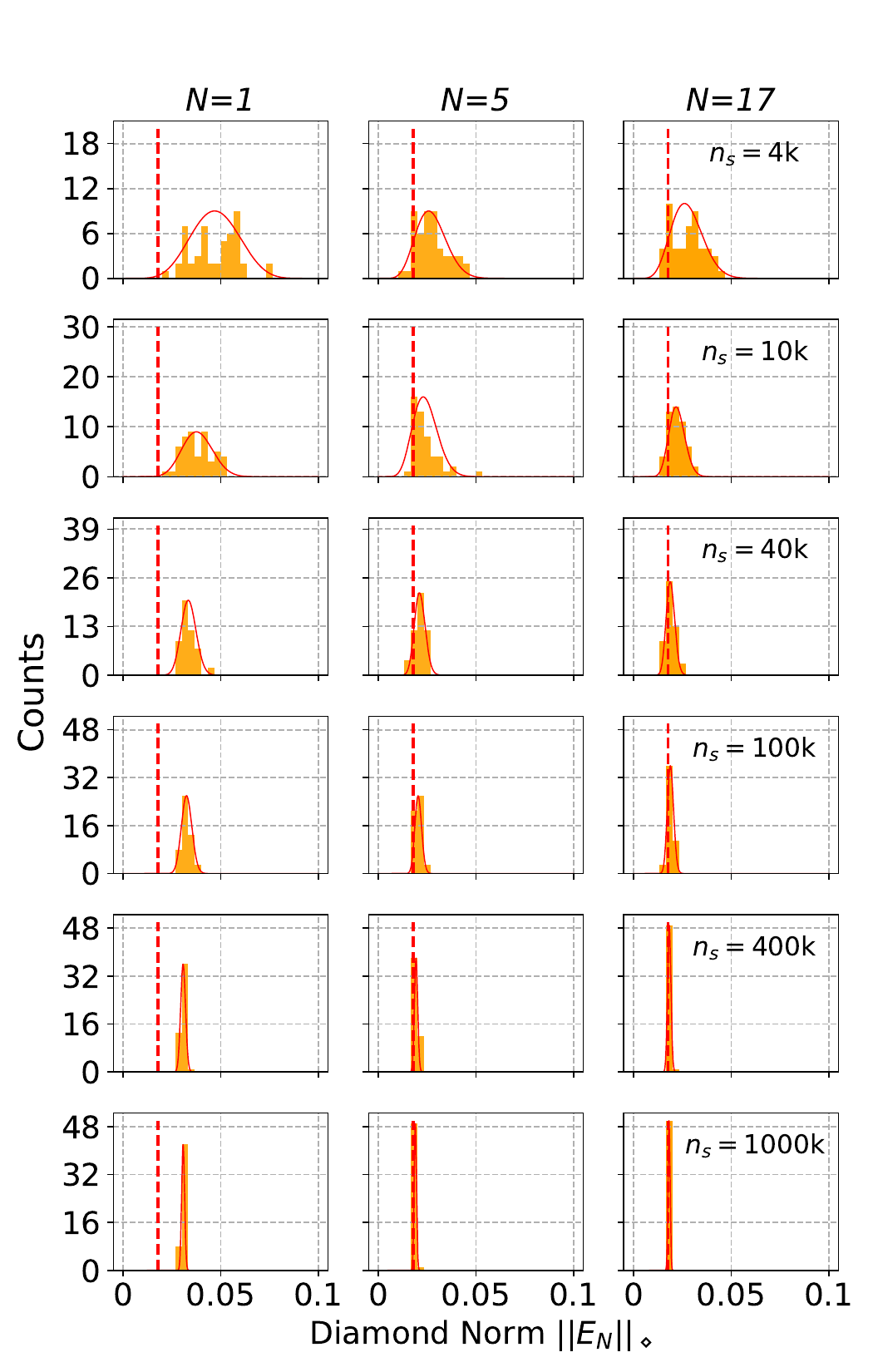}
&
\includegraphics[width=0.9\columnwidth]{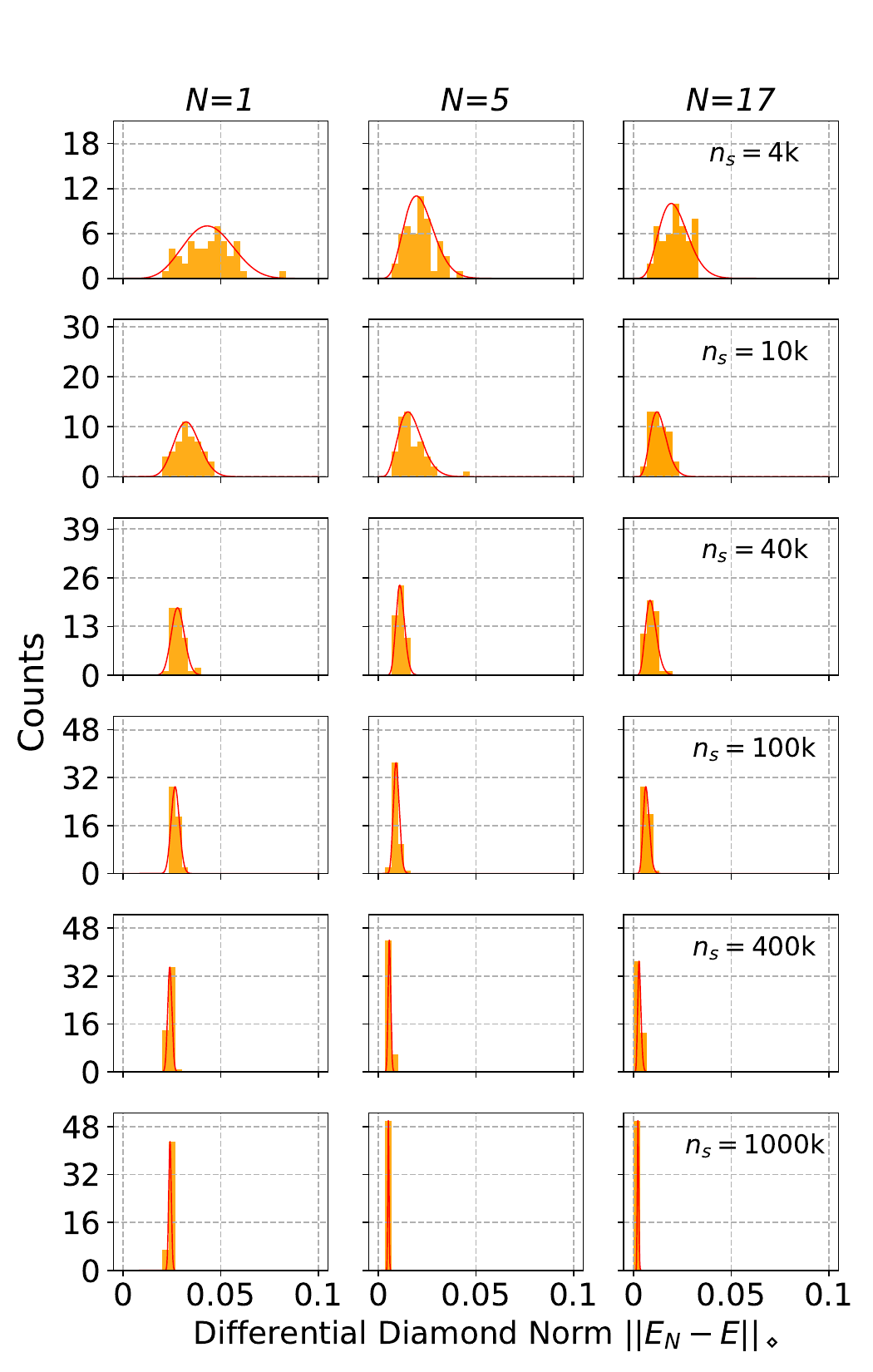}
\end{tabular}
\caption{
Distribution of  the diamond norm $\|\E_N\|_{\diamond}$ (left) and the differential diamond norm $\|\E_N-\E\|_{\diamond}$ (right) for the $\sqrt{X}$ gate for various numbers of passes $N = 4m+1$ and number of shots $n_s$ calculated by using the iterative \MPPT~method. 
The histograms display the results from 50 different simulations,
 in order to obtain reliable statistics, related to the limited number of shots $n_s$.
The solid red curves indicate the beta distribution fit to the data.
The dashed line in the left matrix of plots indicates the actual value of the error norm.
}
     \label{Fig:SX_diam_dist_iter}
\end{figure*}
%%%%%%%%%%%% Fig %%%%%%%%%%%% Fig %%%%%%%%%%%% Fig %%%%%%%%%%%% Fig %%%%%%%%%%%%

In IBM Qiskit, the output of the \textsc{ProcessTomography} measurement is presented by the measured Choi matrix $\mathcal{C}$. 
In the first stage of \MPPT, when performing QPT on $N$ gates we have a measured Choi matrix $\mathcal{C}^{(N)}$, that we convert to the respective PTM $\mathcal{R}^{N}$ by the transformation
\be 
\left(\mathcal{R}^{N}\right)_{i, j}=\frac{1}{d}\operatorname{Tr}\left[\mathcal{C}^{(N)} P_j^T \otimes P_i\right].
\ee
Subsequently, in the post-processing \MPPT~ procedure, having found $\mathcal{R}^{N}$, we determine the measured error matrix $\E_{N}$ and the measured single process $\rm = \T+\E_{N}$. 
Since we know the actual error $\E$, we can visualize the precision and accuracy enhancement by using the diamond norms of the measured error matrix $\|\E_{N}\|_{\diamond}$ and the difference between the measured and actual error matrices $\|\E_{N}-\E\|_{\diamond}$.

\subsection{\MPPT~simulation of $\sqrt{X}$}\label{Dem_sim_sx}

%\subsubsection{Error matrices $\E_N$, $\sqrt{X}$ gate}

%\subsubsection{Diamond norms, $\sqrt{X}$ gate}

We begin with a simulation of the single-qubit gate $\sqrt{X}$, which is a basic gate in IBM Quantum processors. 
The target $\T$ and the actual error matrix $\E$ are plotted in Fig.~\ref{Fig:SX_PTM_E_En1} (top row), and the  explicit form of  $\E$ is given in Appendix \ref{A1}. 
%The error matrix has a diamond norm $\|\E\|_{\diamond}=0.018$ and infidelity $e_F=0.00019$. 
In each shot, we applied the $\sqrt{X}$ gate $N=4m+1$ times ($N$ passes) and then performed standard QPT in order to generate the PTM matrix of the process.
Because each shot has a statistical shot noise error, we used $n_s$ shots in order to reduce this error. 
In the simulations we used $n_s=$ 4k to 1000k. 
For each combination of \(N\) and \(n_s\), we executed 50 tomographies, followed by the application of the iterative method with some comparisons with the linear one.
Each of the 50 tomographies had its own ``shot noise'' statistical error and hence their distribution is presented in histograms.

Figure~\ref{Fig:SX_PTM_E_En1} provides a visual representation of the enhanced \MPPT~by displaying the error matrices $\E_N$ for various numbers of passes ($N=1,5,17$) and numbers of shots (10k, 100k, 1000k). 
When compared to the actual matrix $\E$, as shown in Fig.~\ref{Fig:SX_PTM_E_En1} (top/right), it becomes evident how the matrices $\E_N$ progressively approximate $\E$ more closely with the increase in both $N$ and $n_s$, 
while the single-pass matrices $\E_1$ differ significantly, even for 1000k shots.
%from the actual matrix $\E$ shown on Fig.~\ref{Fig:SX_PTM_T_E}(right), while the matrices for increasing numbers of passes $N$ and shots approach closer to  $\E$. 

In order to conduct a more thorough examination of the precision and accuracy of our method, we analyzed the distributions of the diamond norms $\|\E_N\|_{\diamond}$ and $\|\E_N-\E\|_{\diamond}$ derived by the iterative method, which are depicted in Fig.~\ref{Fig:SX_diam_dist_iter}.
%and \ref{Fig:SX_dDm_dist_iter}. 
The left-hand-side matrix of plots reveals that for SQPT ($N=1$) only the precision is improved as the number of shots $n_s$ increases, while the accuracy is still low. 
The reason is that the SPAM and readout errors add up to the gate errors and prevent SQPT from retrieving the latter.
On the contrary, for \MPPT~($N=5, 17$), the distributions squeeze (indicating improved precision) and shift closer to the exact values (indicating improved accuracy) when $N$ and $n_s$ increase.
%indicating enhancements in both precision and accuracy.

%These distributions show that with the application of \MPPT, there is a noticeable improvement in both precision and accuracy, with the \MPPT~distributions converging toward the actual diamond norm value $\|\E\|_{\diamond}$ (depicted by the red dashed line). For SQPT ( $N=1$), while precision is enhanced as  $n_s$ increases, the accuracy does not exhibit a corresponding increase. The solid red line delineates the beta distribution fit for the data.

The right-hand-side matrix of plots in Fig.~\ref{Fig:SX_diam_dist_iter} shows the distribution of the diamond norm difference, $\|\E_N-\E\|_{\diamond}$, representing the distance between the measured error matrix $\E_N$ and the actual error matrix $\E$, for various numbers of passes $N = 4m+1$ and numbers of shots $n_s$. 
This figure demonstrates how effectively MQPT can measure the actual matrix $\E$, going beyond merely assessing the actual diamond norm $\|\E\|_{\diamond}$, as compared to Fig.~\ref{Fig:SX_diam_dist_iter}. 
Indeed, different matrices $\E_N$ can have the same diamond norm $\|\E_N\|_{\diamond}$ but they can be at different distances  $\|\E_N-\E\|_{\diamond}$ from the actual error matrix $\E$.
However, when the diamond distance $\|\E_N-\E\|_{\diamond}$ approaches 0 then the calculated error matrix $\E_N$ approaches the actual error matrix  $\E$ with certainty.
The left column in the right-hand-side matrix of plots in Fig.~\ref{Fig:SX_diam_dist_iter} demonstrates that this cannot be achieved with a single pass regardless of the number of shots.
The multipass \MPPT~method, even for the relatively low values of $N$ in the figure, make it possible to approach very closely the actual error matrix $\E$ because the detrimental SPAM and readout effects are greatly reduced.

%%%%%%%%%%%% Fig %%%%%%%%%%%% Fig %%%%%%%%%%%% Fig %%%%%%%%%%%% Fig %%%%%%%%%%%%
\begin{figure}[tb]
\begin{tabular}{cc}
\includegraphics[width=\columnwidth]{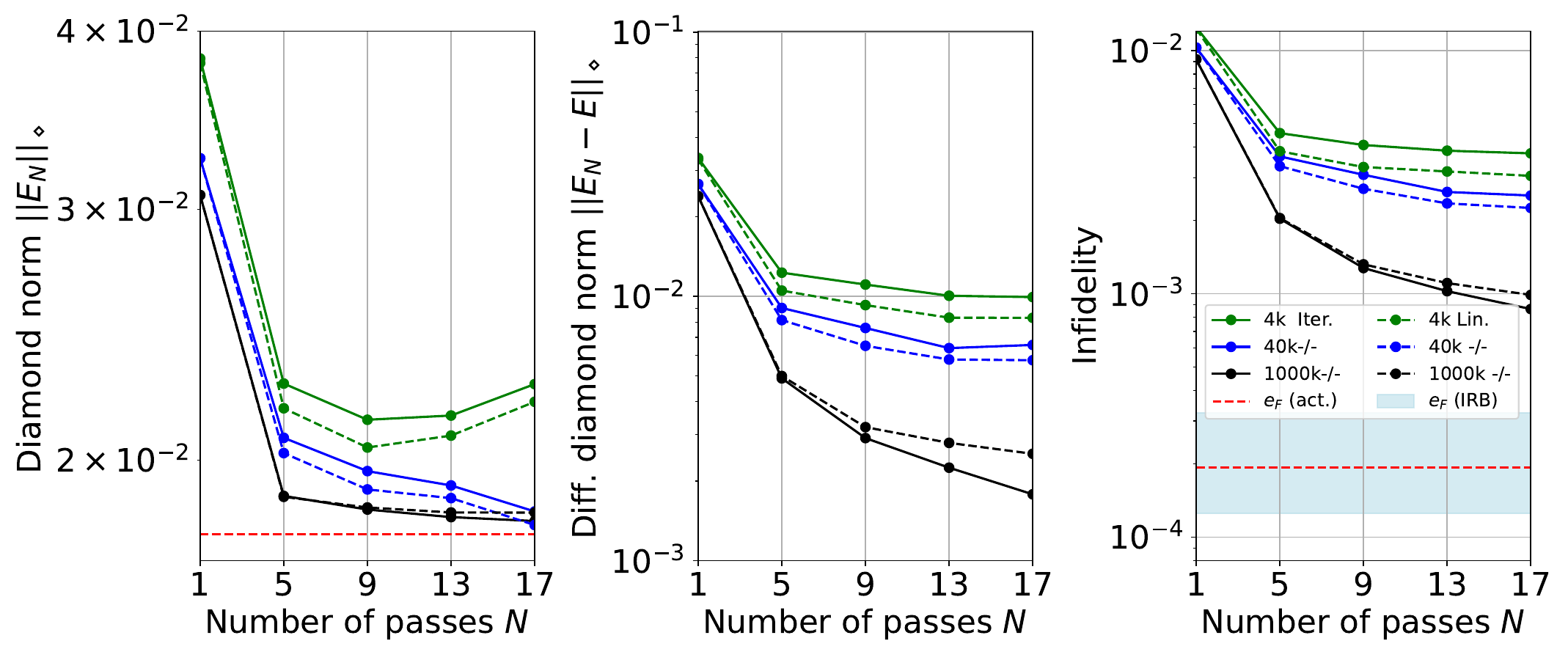}
\end{tabular}
\caption{Comparison of the iterative \MPPT~(solid lines), linear \MPPT~(dashed lines) and Interleaved Randomized Benchmarking (IRB, band) methods  applied to the $\sqrt{X}$ gate. 
We compare the diamond norm $\|\E_N\|_{\diamond}$ (left), the differential diamond norm $\|\E_N - \E\|_{\diamond}$ (middle) and the infidelity (right) for several numbers of passes $N$ and several numbers of shots $n_s$. 
The actual infidelity $e_F$ is shown by the horizontal dashed line in the left and right plots and serves as the target value.
The target value in the middle plot is 0.
%\textcolor{red}{\sc NV: Zashto sa v italic zaglaviqta na osite?}
}
     \label{Fig:SX_diam_inf}
\end{figure}
%%%%%%%%%%%% Fig %%%%%%%%%%%% Fig %%%%%%%%%%%% Fig %%%%%%%%%%%% Fig %%%%%%%%%%%% 

While distribution plots for the linear method have been omitted due to their qualitative resemblance to those of the iterative method, Fig.~\ref{Fig:SX_diam_inf} presents a comparison of the two methods. %aiding in assessing their relative accuracy. 
It is noteworthy that, for a range of shots (4k and 40k), both $\|\E_N\|_{\diamond}$ and $\|\E_N-\E\|_{\diamond}$ yield smaller values by the linear method. 
This observation seems to contradict expectations, as the iterative method typically offers a practically exact solution for finding $\E$ given a multi-pass process $\M$. 
However, the linear method mixes the post-processing with the shot noise errors, which statistically leads to a random error compensation.
This implicit feature is quite similar to Ridge regression~\cite{Gruber1998}, which prevents overfitting to noisy data, and to the Wiener filter~\cite{Boulfelfel1994}, suggesting that the linear method might provide a more stable solution in the presence of noise. 
However, this trend is reversed for the higher shot count of 1000k, 
%where the iterative method demonstrates higher accuracy than the linear approach. 
%Upon further increase of the number of shots,
making the data clearer of shot noise error, the inherent ability of the iterative method to precisely adjust to the signal without being distorted by noise comes to the forefront. 
At the same time, the linear method begins to exhibit its limitations, as at reduced levels of shot noise, an increasingly uncompensated post-processing error becomes evident, and the accuracy of the linear method falls behind.

%From 4k to 100k shots, the linear method yields lower norms than the iterative method, which shows that the linear method works as a shot noise filter similar to Wiener filter. However, as the shot count reach 400k, the trend reverses, with iterative norms becoming smaller than the linear ones. This suggests that at higher shot counts, the iterative method more effectively mitigates the SPAM and readout errors. 

%\subsubsection{Infidelity, $\sqrt{X}$  gate. Comparison with IRB}

%%%%%%%%%%%% Fig %%%%%%%%%%%% Fig %%%%%%%%%%%% Fig %%%%%%%%%%%% Fig %%%%%%%%%%%%
\begin{figure}[tb]
    \centering
    \includegraphics[width=\columnwidth]{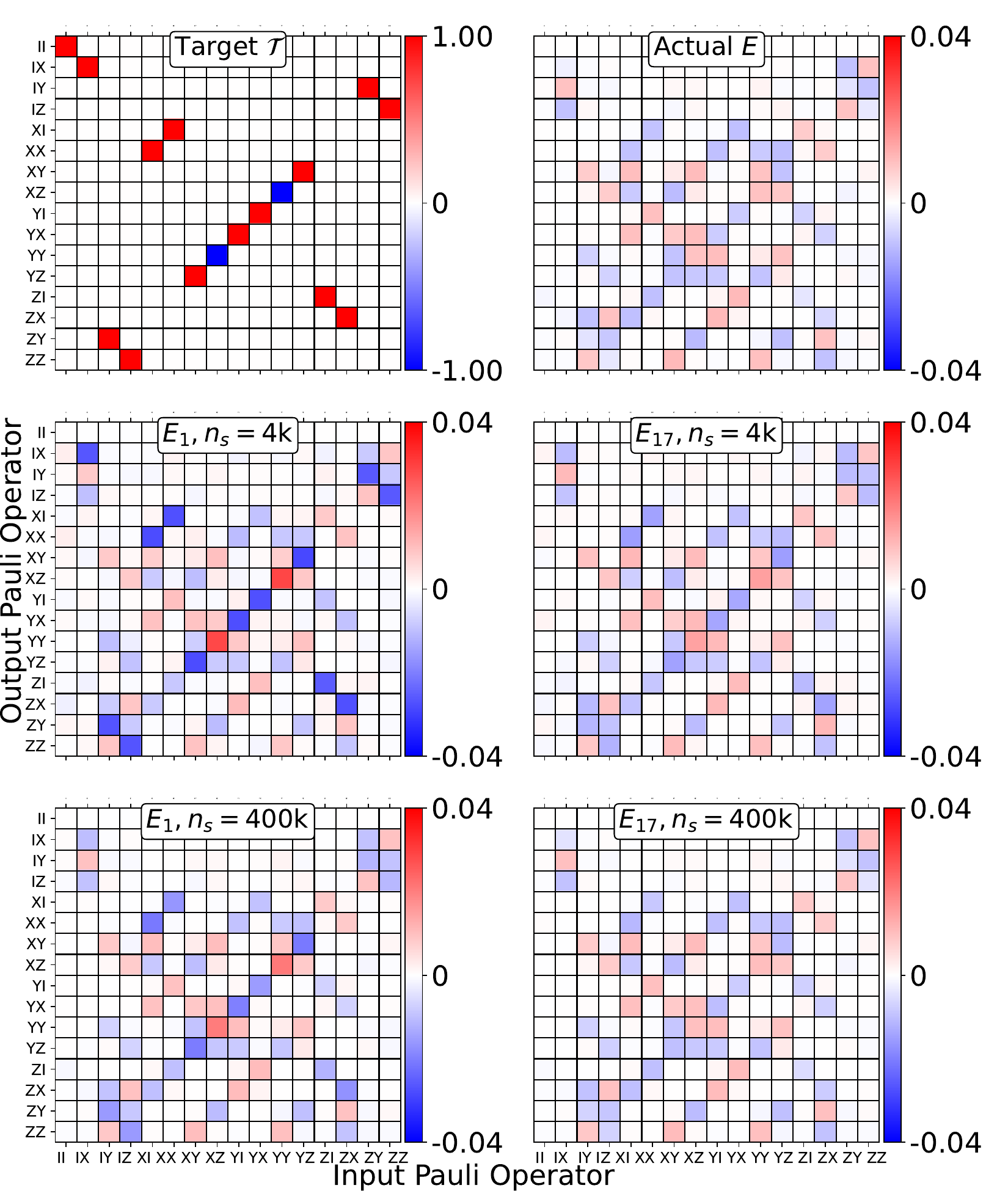}
    \caption{Target CNOT Pauli Transfer matrix $\T$ (left) and the actual error matrix $\E$ (right),  which is randomly generated for our simulations with process infidelity $e_{F}=0.0062$ and diamond norm $\|\E\|_{\diamond}=0.073$. The actual process $\R$ is the sum of these two matrices. On this channel, we simulate SQPT, by using of IBM \textsc{ibmq\_qasm\_simulator} and the Qiskit \textsc{ProcessTomography} module. 
    }
    \label{fig:CNOT_PTM_E_Ns}
\end{figure}

%%%%%%%%%%%% Fig %%%%%%%%%%%% Fig %%%%%%%%%%%% Fig %%%%%%%%%%%% Fig %%%%%%%%%%%%
  
%%%%%%%%%%%% Fig %%%%%%%%%%%% Fig %%%%%%%%%%%% Fig %%%%%%%%%%%% Fig %%%%%%%%%%%%
\begin{figure*}[tb]
\begin{tabular}{cc}
\includegraphics[width=0.9\columnwidth]{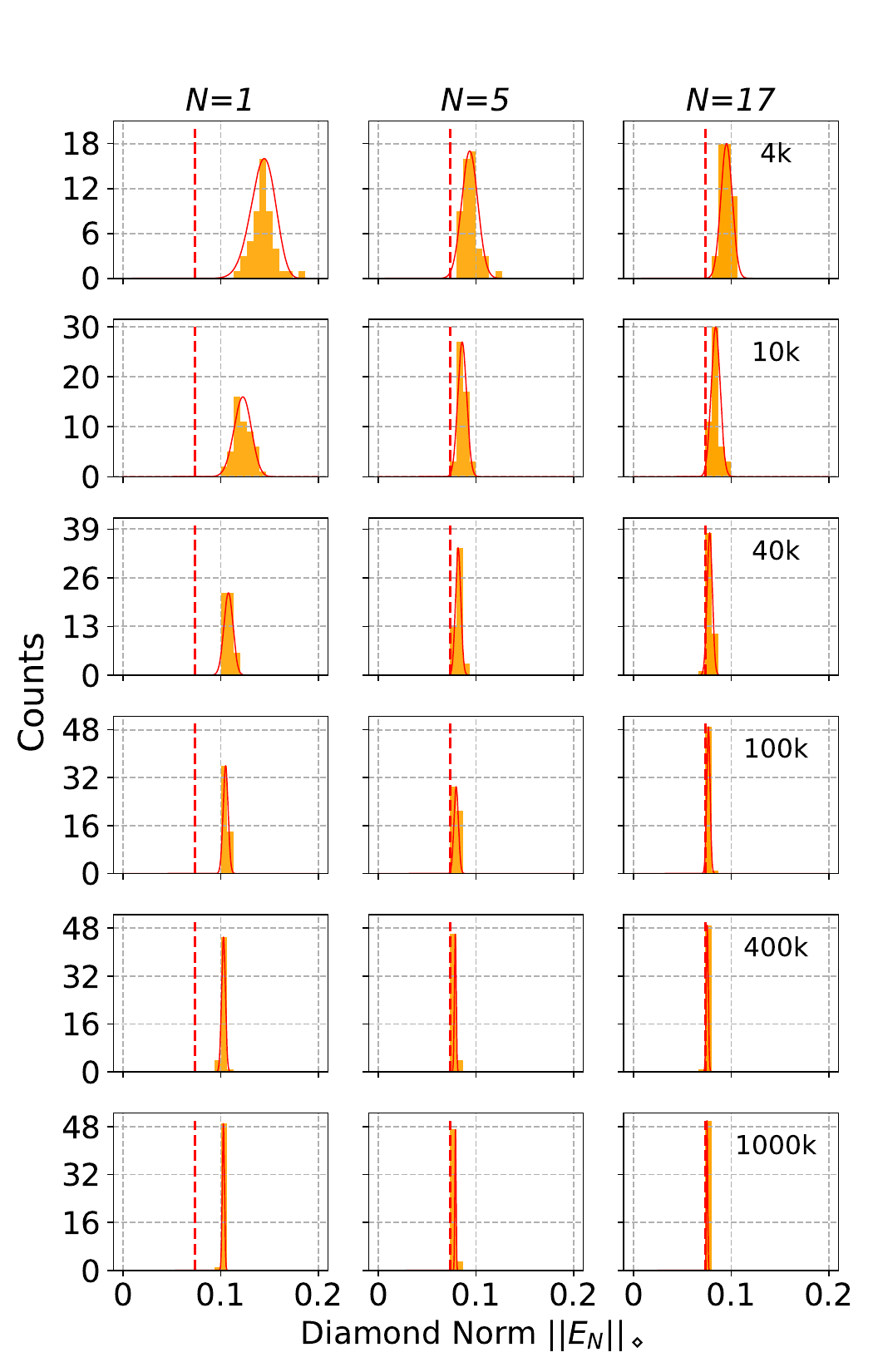}
&
\includegraphics[width=0.9\columnwidth]{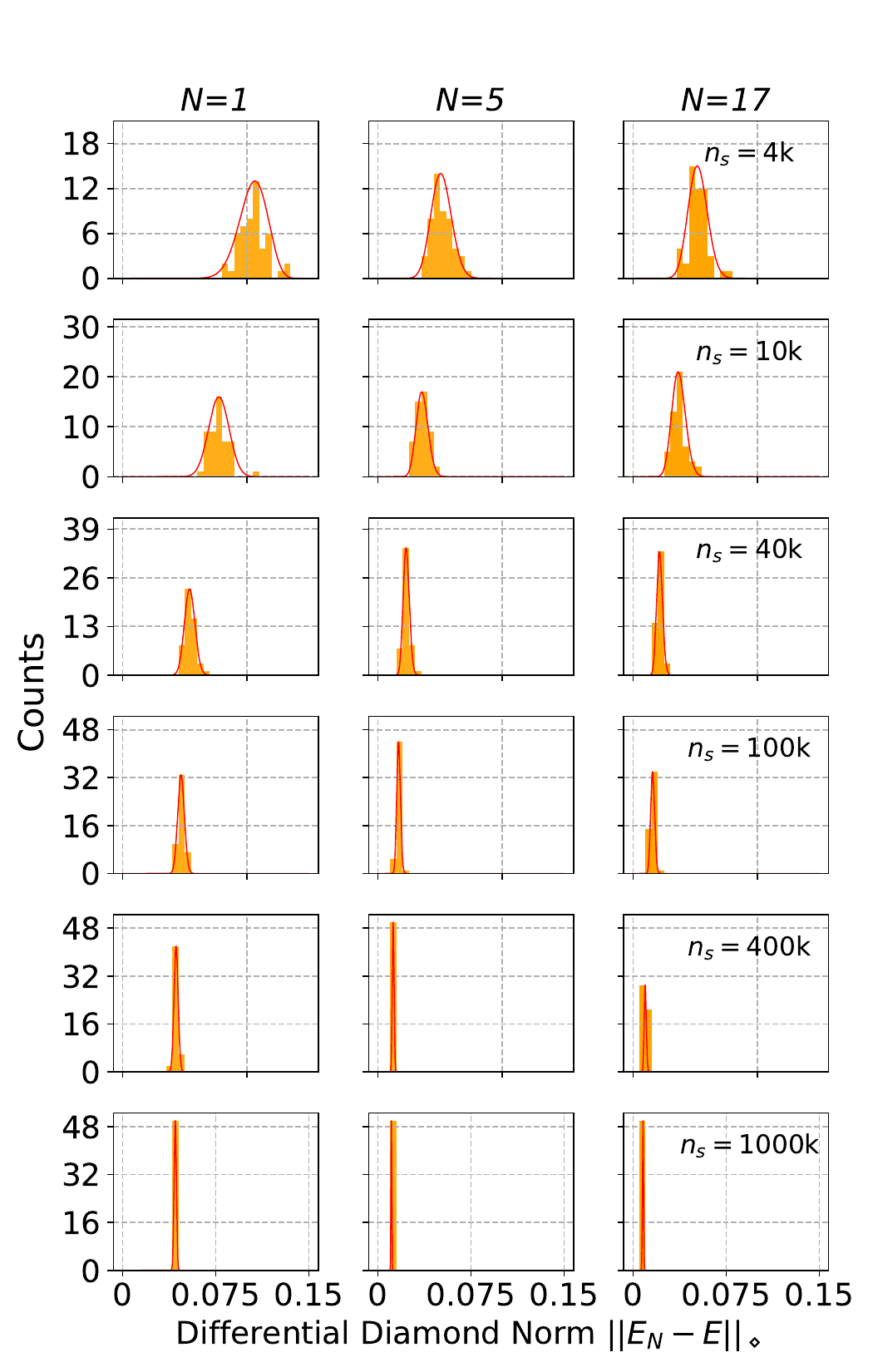}
\end{tabular}
\caption{
The same as Fig.~\ref{Fig:SX_diam_dist_iter} but for the CNOT gate.
%Distribution of  the diamond norm $\|\E_N\|_{\diamond}$ (left) and the differential diamond norm $\|\E_N-\E\|_{\diamond}$ (right) for the CNOT gate  for various numbers of passes $N = 4m+1$ and number of shots $n_s$ calculated by using the iterative \MPPT~method. 
%The histograms display the results from 50 different experiments.
%\textcolor{red}{\sc NV: What is different?}
%The solid red curves indicate the beta distribution fit to the data.
%The dashed line in the left matrix of plots indicates the actual value of the error norm.
%
}
     \label{Fig:CNOT_diam_dist_iter}
\end{figure*}
%%%%%%%%%%%% Fig %%%%%%%%%%%% Fig %%%%%%%%%%%% Fig %%%%%%%%%%%% Fig %%%%%%%%%%%%

In order to showcase the accuracy of infidelity measurement by \MPPT, we analyzed results from both linear and iterative methods, alongside an infidelity measurement conducted via the IRB method using the available Qiskit module \textsc{InterleavedRB}. 
Figure~\ref{Fig:SX_diam_inf} (right) displays these results on a logarithmic scale. 
Echoing our earlier observations in the diamond norm comparison, the linear method demonstrates greater accuracy in high shot noise (small number of shots) situations. 
\MPPT, by contrast, attains comparable accuracy levels when a large number of shots (exceeding 400k) and multiple passes $N$ are implemented. 
Another noteworthy observation is the precise infidelity measurement achieved by the IRB method. 
While the IRB method accurately determines infidelity $e_{F}$, it does not provide information about the error matrix $\E$ and the differential diamond norm $\|\E_N-\E\|_{\diamond}$. 
In contrast, the strength of \MPPT~lies in its ability to deliver these metrics, although with lower accuracy in the measurement of $e_{F}$ compared to IRB.

%showing that IRB provides a relatively accurate measure $e_F(IRB)$. 
%Achieving comparable accuracy with MQPT requires a high shot volume ($n_s > 400\)k) and multiple passes ($N$). 
%While IRB offers a straightforward and precise assessment of $e_F$, it does not yield information about the actual error matrix $\E$, a gap that MQPT can fill in.

\subsection{\MPPT~simulation of CNOT gate}\label{Dem_sim_cnot}

Following the same \MPPT~procedure as for $\sqrt{X}$ gate, in this section we apply it to CNOT gate. 
We use an imperfect CNOT gate with a target PTM $\T$ and error matrix $\E$ shown in Fig.~\ref{fig:CNOT_PTM_E_Ns} (top right).
It has a diamond norm $\|\E\|_{\diamond} = 0.073$, infidelity of $e_{F}=0.0062$ and its explicit form is shown in Appendix~\ref{A1}.      
The CNOT gate that is characterized in the simulation has the most significant bit (MSB) as the control qubit, i.e. cx(1,0). 
According to the Qiskit MSB convention, the corresponding $4 \times 4$ target gate is represented as  
\be
\bm{T}_{cnot} \equiv
\begin{quantikz}
    \lstick{$q_0$} & \targ{1} & \qw \\
    \lstick{$q_1$} & [short] \ctrl{-1} & \qw  
\end{quantikz}
\equiv \left(\begin{array}{cccc}
1&0&0&0\\
0&1&0&0\\
0&0&0&1\\
0&0&1&0
\end{array}\right) \equiv \text{cx}(1,0) .
\ee

%\subsubsection{Error matrices $\E_N$, CNOT gate}
Figure~\ref{fig:CNOT_PTM_E_Ns} (middle and bottom rows) provides a visual representation of the enhanced \MPPT~by displaying the error matrices $\E_N$ for 1 and 17 passes and $n_s=4$k and $400$k shots. 
The matrices are averaged over 50 individual tomographies. 
%SThrough these tools, we explore how \MPPT~enhances the precision and accuracy for various noise models that simulate SPAM, readout, and shot noise errors.
Compared to the actual error matrix $\E$, as shown in Figure~\ref{fig:CNOT_PTM_E_Ns} (top right), it becomes evident how the \MPPT~matrices $\E_{17}$ approximate $\E$ more closely than the SQPT matrices $\E_{1}$.
Moreover, the plots show that increasing the number of passes $N$ brings the multipass simulated error matrix $\E_{17}$, with a small number of shots (4k) much closer to the actual error matrix $\E$ than what the simulated single-pass error matrix $\E_1$ can achieve, even with a much larger number of shots $n_s=400$k.

Following the approach taken for the $\sqrt{X}$ gate, we examine the diamond norm distributions for the CNOT gate in order to assess the precision and accuracy of our \MPPT~approach. 
The diamond norms $\|\E_N\|_{\diamond}$ and $\|\E_N-\E\|_{\diamond}$, calculated with the iterative method, are plotted in Fig.~\ref{Fig:CNOT_diam_dist_iter}.
As for the $\sqrt X$ gate, the right-hand-side matrix of plots, which shows the difference norm $\|\E_N-\E\|_{\diamond}$ between the simulated $\E_N$ and actual $\E$ error matrices, is the unambiguous measure of success or failure of the simulation, because different simulated matrices $\E_N$ may have the same diamond norm $\|\E_N\|_{\diamond}$ but they may be far from the actual error matrix $\E$.

Figure \ref{Fig:CNOT_diam_dist_iter} demonstrates that, as the number of passes $N$ increases, the distribution becomes both more precise (its width shrinks) and more accurate (it approaches the vertical dashed line, which shows the actual error stemming from the generated random error matrix).
On the other hand, when we increase the number of shots $n_s$ and hence reduce the statistical shot-noise error, the precision increases (according to the statistical square-root law), but the change in accuracy depends on the number of passes used: it quickly saturates for 1 pass, while for 5 and 17 passes the distribution moves steadily toward the dashed line (the actual error).

The reason is that the SPAM and readout errors shift the single-pass simulations, whereas these errors are greatly reduced by the multipass approach.
This simulation clearly conveys the enhancements in both precision and accuracy with the \MPPT~repetition method.

%%%%%%%%%%%% Fig %%%%%%%%%%%% Fig %%%%%%%%%%%% Fig %%%%%%%%%%%% Fig %%%%%%%%%%%%
\begin{figure}[tb]
\begin{tabular}{cc}
\includegraphics[width=\columnwidth]{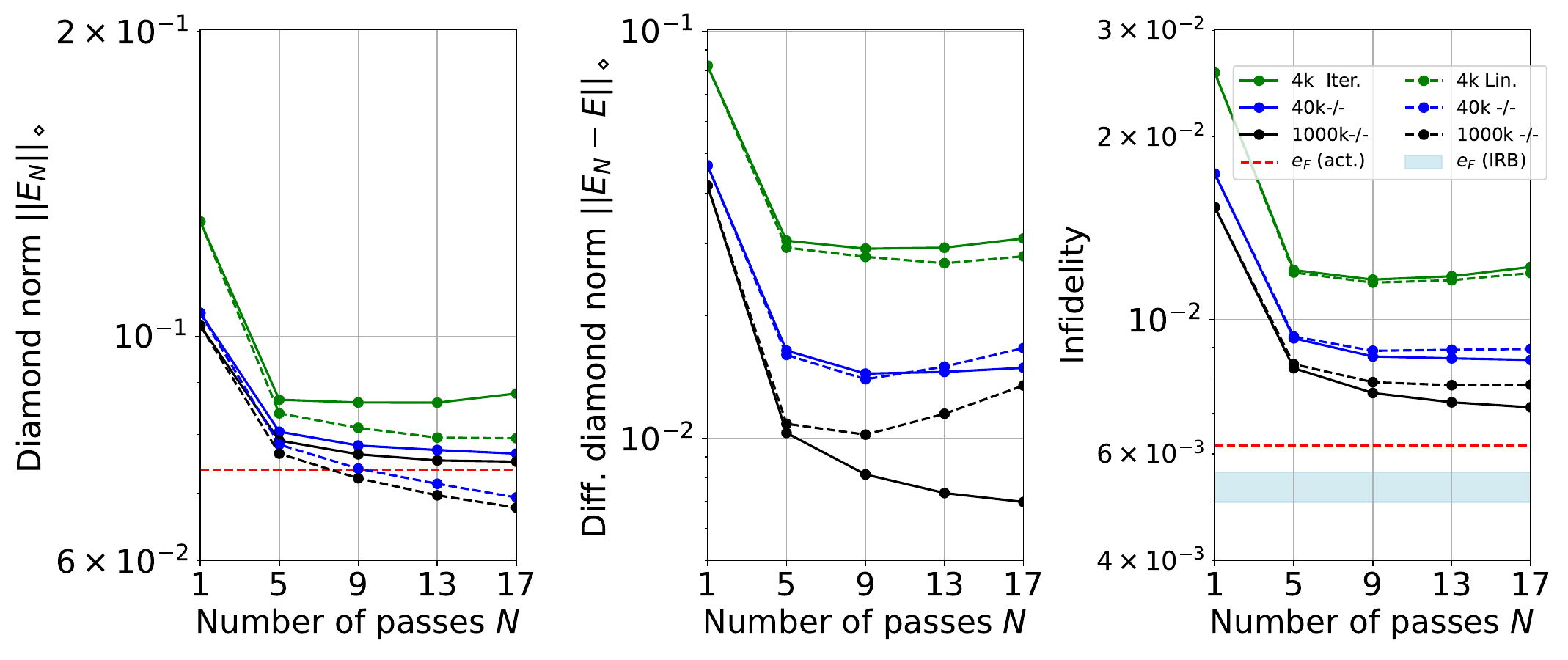}
\end{tabular}
\caption{
The same as Fig.~\ref{Fig:SX_diam_inf} but for the CNOT gate. %\textcolor{blue}
{The IRB value is slightly lower than the actual value because IRB gives the averaged process fidelity, instead of process fidelity, that is more rigorous and utilized in this paper.}}

     \label{Fig:CNOT_diam_iter_lin}
\end{figure}
%%%%%%%%%%%% Fig %%%%%%%%%%%% Fig %%%%%%%%%%%% Fig %%%%%%%%%%%% Fig %%%%%%%%%%%%

Fig.~\ref{Fig:CNOT_diam_iter_lin} presents a comparison of the diamond norms of the CNOT gate calculated by the linear and iterative \MPPT~methods.
It is a test of how the two methods perform relative to each other.
For 4k shots, the linear method yields better accuracy than the iterative method. 
However, as the shot count exceeds 40k, the trend reverses, with the iterative norms becoming the more accurate ones. 
This suggests that at higher shot counts, the iterative method more effectively mitigates the SPAM and readout errors. 
Additionally, at 40k and 1000k shots, there is a monotonic decrease in the iterative norms as the number of passes $N$ increases, further emphasizing the iterative method's advantage in the high-shot regime.

It is important to note that the diamond norm $\|\E_N\|_{\diamond}$ calculated by the linear method for 40k and 1000k shots crosses the line of the actual norm $\E$ for more than 9 passes, i.e. it underestimates the error (left plot). 
Conversely, the iterative method asymptotically approaches the actual norm $\E$ as the number of passes $N$ increases. 
This observation does not show up in the differential diamond norm $\|\E_N-\E\|_{\diamond}$ (middle plot), which is a better measure of the methods performance.

%\subsubsection{Infidelity, CNOT  gate. Comparison with IRB}

The situation with the simulated infidelity is similar as for the $\sqrt{X}$ gate. 
Figure~\ref{Fig:CNOT_diam_iter_lin} (right) displays the infidelity of the CNOT gate where the exact infidelity is compared to the infidelities calculated by the linear and iterative \MPPT~methods and the IRB method.
The infidelity simulated by IRB is more accurate than \MPPT~for small numbers of passes $N$ and shots $n_s$, but for larger values of $N$ and $n_s$ the \MPPT~method features similar accuracy as IRB. 
%The linear \MPPT~method demonstrates a bit greater accuracy in high shot noise $n_s=4k$. 
%, showing that IRB provides a relatively accurate measure $e_F(IRB)$. 
%Achieving comparable accuracy with MQPT requires a high shot volume ($n_s > 400\)k) and multiple passes ($N$). 
While IRB offers a straightforward and fairly accurate assessment of $e_F$, it only gives a single number and  does not yield the far more extensive information about the actual error matrix $\E$: this is what MQPT~delivers.

%\textcolor{red}

An interesting observation regarding IRB follows from comparing Figs.~\ref{Fig:SX_diam_inf} and \ref{Fig:CNOT_diam_iter_lin}.
While the IRB method correctly finds the exact infidelity for the $\sqrt X$ gate in Fig.~\ref{Fig:SX_diam_inf}, it fails to do so for the CNOT gate in Fig.~\ref{Fig:CNOT_diam_iter_lin}.
We have found that this feature is not specific for the particular random choice of the actual error matrix $\E$ but holds in general.

%%%%%%%%%%%%%% SEC %%%%%%%%%%%%%% SEC %%%%%%%%%%%%%% SEC %%%%%%%%%%%%%% SEC %%%%%%%%%%%%%% 
%%%%%%%%%%%%%% SEC %%%%%%%%%%%%%% SEC %%%%%%%%%%%%%% SEC %%%%%%%%%%%%%% SEC %%%%%%%%%%%%%% 
%%%%%%%%%%%%%% SEC %%%%%%%%%%%%%% SEC %%%%%%%%%%%%%% SEC %%%%%%%%%%%%%% SEC %%%%%%%%%%%%%% 
\section{Experimental demonstration of \MPPT~on \textsc{ibmq\_manila}}\label{ExpDemManila}

%%%%%%%%%%%% Fig %%%%%%%%%%%% Fig %%%%%%%%%%%% Fig %%%%%%%%%%%% Fig %%%%%%%%%%%%
\begin{figure}[tb]
    \makebox[\linewidth]{
\includegraphics[width=\columnwidth]{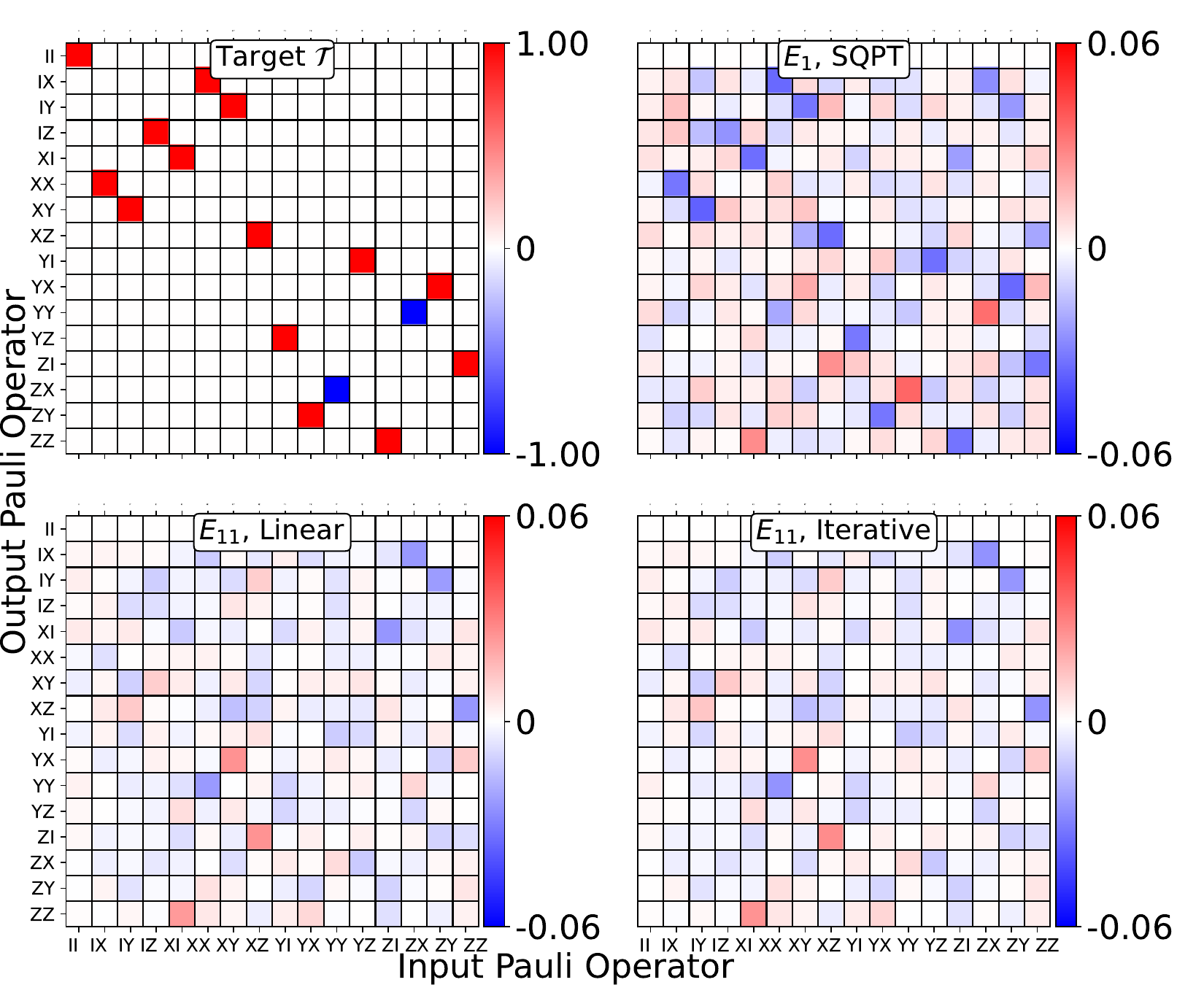}
    }
\caption{Top/left - CNOT/cx(0,1)/ PTM target matrix  ; Top/right - Measured error matrix  $\E_1$ by SQPT ($N=1$); bottom/left - \MPPT~ $\E_{11}$  with the linear method; bottom/right - \MPPT~ $\E_{11}$  with the iterative method.
All error matrices are averaged over ten tomographies performed with the Qiskit module \textsc{MitigatedProcessTomography}. }
     \label{Fig:PTM_T_E_Manila}
\end{figure}
%%%%%%%%%%%% Fig %%%%%%%%%%%% Fig %%%%%%%%%%%% Fig %%%%%%%%%%%% Fig %%%%%%%%%%%%

In this section, we demonstrate \MPPT~experimentally on the CNOT gate that is the basic two-qubit gate on \textsc{ibmq\_manila} processor ~\cite{IBM} and we compare it with the results of SQPT.  
In order to improve the quality of SQPT, we use the Qiskit module \textsc{MitigatedProcessTomography}, that mitigates the readout error. 
This module includes two additional circuits to the regular 144 SQPT circuits that are in the module \textsc{ProcessTomography}. 
We performed SQPT with $N=1$ pass and \MPPT~with $N=11$ passes. 
For both SQPT and \MPPT~we conducted 10 tomographies with $n_s=4$k shots.  
The CNOT that is characterized in the experimental demonstration has a least significant bit (LSB) as the control qubit, i.e. cx(0,1). 
According to the IBM Qiskit LSB convention, the corresponding $4 \times 4$ target gate is represented by
\begin{equation}
T_{cnot} \equiv
\begin{quantikz}
    \lstick{$q_0$} & \ctrl{1} & \qw \\
    \lstick{$q_1$} & [short] \targ{} & \qw  
\end{quantikz}
\equiv \left(\begin{array}{cccc}
1&0&0&0\\
0&0&0&1\\
0&0&1&0\\
0&1&0&0
\end{array}\right) \equiv \text{cx}(0,1)
\end{equation}

The respective PTM target matrix is shown in Fig.~\ref{Fig:PTM_T_E_Manila} (top left),
and the SQPT error matrix  $\E_1$ in  Fig.~\ref{Fig:PTM_T_E_Manila} (top right). 
On the bottom row of  Fig.~\ref{Fig:PTM_T_E_Manila} we present the measured error matrix $\E_{11}$ by \MPPT~with the linear and iterative methods. 
Both SQPT and \MPPT~ matrices are averaged over 10 tomographies. 
The target CNOT matrix $\T$ and the measured error matrix $\E_{11}$ are given explicitly in Appendix A.
As in the simulations in the preceding section, SQPT overestimates the errors (due the uncompensated SPAM and readout errors), while both \MPPT~methods give significantly lower (and similar to each other) values.

%\subsection{Infidelity measurement results }

%%%%%%%%%%%% Fig %%%%%%%%%%%% Fig %%%%%%%%%%%% Fig %%%%%%%%%%%% Fig %%%%%%%%%%%%
\begin{figure}[tb]
    \makebox[\linewidth]{
\includegraphics[width=0.7\columnwidth]{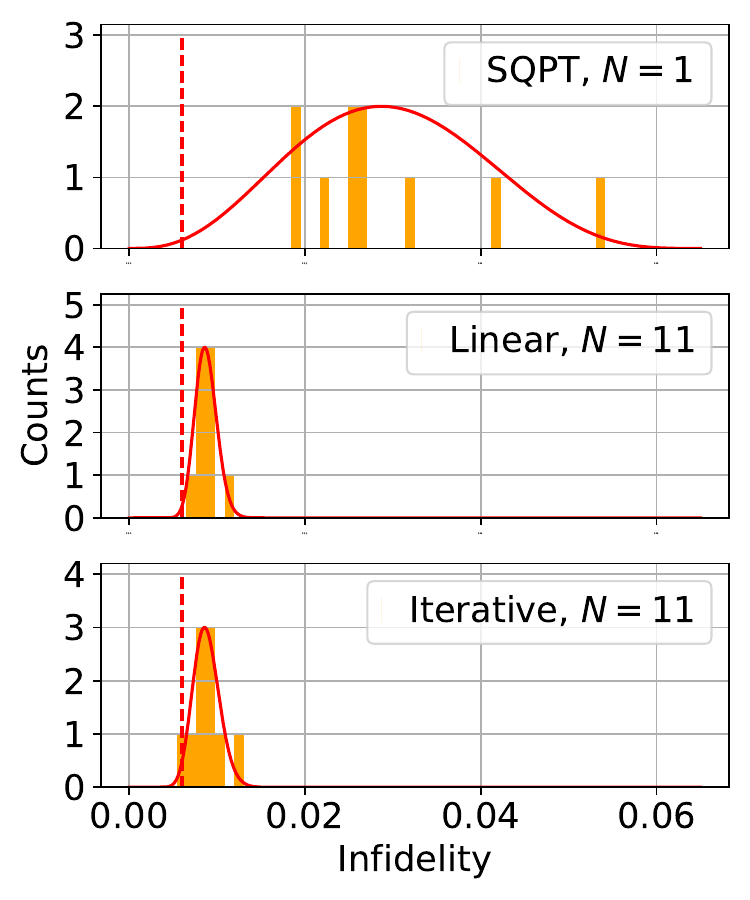}
    }
\caption{CNOT Infidelity distribution after SQPT and \MPPT~ (linear and itearive) with $N=11$ on IBM Quantum Device \textsc{ibmq\_manila}.
Ten experiments were conducted using the Qiskit module \textsc{MitigatedProcessTomography}, each with $n_s=4$k shots. The median infidelity is taken from the IBM Quantum Calibration data (red dashed). The Beta curves (red doted) represent the fitted infidelity distributions for SQPT and \MPPT. 
}
     \label{Fig:infid_Manila}
\end{figure}
%%%%%%%%%%%% Fig %%%%%%%%%%%% Fig %%%%%%%%%%%% Fig %%%%%%%%%%%% Fig %%%%%%%%%%%%

In  Fig.~\ref{Fig:infid_Manila} we show the results for the infidelity distribution performed with 10 tomographies for SQPT and \MPPT~(linear and iterative methods). 
As references for the measurement, we use the IBM Calibration data that specify a \text{CNOT} median gate error of $6\times10^{-3}$ marked with red dashed line. 
We notice significant precision and accuracy enhancements, since the \MPPT~infidelity distribution is much closer to the specified median infidelity (in red dashed) than SQPT. 
%The conclusion we draw from this figure is first, the similarity between the linear and iterative \MPPT~tomographies, and second, the significant difference from SQPT.
With further increase of the number of passes and shots, which unfortunately were not possible, we would expect both the accuracy and the precision to increase.

%\subsection{Direct check via probabilities}

%%%%%%%%%%%% Fig %%%%%%%%%%%% Fig %%%%%%%%%%%% Fig %%%%%%%%%%%% Fig %%%%%%%%%%%%
\begin{figure}[tb]
    \makebox[\linewidth]{
\includegraphics[width=\columnwidth]{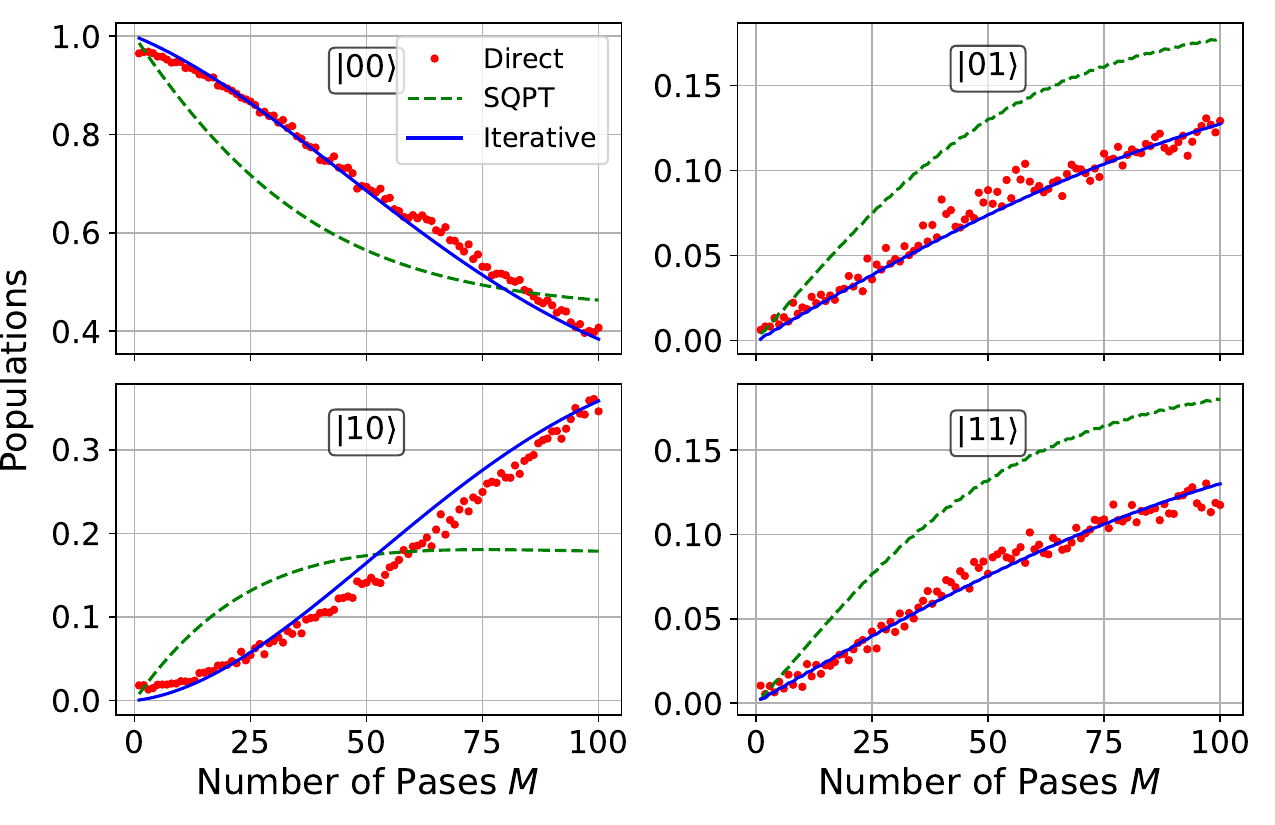}
    }
\caption{Direct/ Indirect probability comparison for SQPT and \MPPT (Linear and  methods) - Experimental demonstration on IBM quantum \textsc{ibmq\_manila}. The direct basic state probabilities (doted marks) are measured after applying of CNOT gate  $M$ times on the initial state $|00\rangle$ at $n_s=4$k. The indirect probabilities (dashed lines) are calculated by the averaged  SQPT and \MPPT~ Liouville processes $\L_1$ ($N=1)$ and $\L_{11}$ ($N=11$) . This process is then raised to the power of various $M$ and applied to the vectorized initial state $|\rho_0\rangle\rangle = |00\rangle$, i.e. the probabilities are calculated by the vector $\L_{N}^M|00\rangle\rangle$ according to Eq.~\eqref{Manila_probs_M}.
%\textbf{NV: only iterative MQPT - with solid lines.}
}
     \label{Fig:Manila_pops_M}
\end{figure}
%%%%%%%%%%%% Fig %%%%%%%%%%%% Fig %%%%%%%%%%%% Fig %%%%%%%%%%%% Fig %%%%%%%%%%%%

We have also used for comparison some other independent and \emph{directly} measurable data, which do not rely on the median fidelity from the IBM Calibration data. 
To this end, we have conducted a series of measurements of the populations $p_{ij}(M)$ on the four basis states $\ket{00},\ket{01},\ket{10},\ket{11} $, by applying the same CNOT gates multiple times $(M=1,2,\ldots,100)$. 
These directly measured probabilities are depicted in Fig.~\ref{Fig:Manila_pops_M}, marked with dots. 
This measurement can be compared with the results from our measured processes $\R_{N}$ for SQPT ($N=1$)  and \MPPT~($N=11$). 
In order to demonstrate the \emph{indirect} results, we convert the measured Pauli transfer matrices to Liouville matrices, i.e. $\R_{N} \to \L_{N}$. 
We then exponentiate these matrices to various powers of $M$ and we apply them to the initial vectorized (by column stacking) state $|\rho_0\rangle \rangle=|00\rangle \rangle$. 
The indirectly measured probabilities $p_{ij}(N,M)$ are then obtained as 
\bse\label{Manila_probs_M}
\begin{align}    
p_{00}(N,M)&=\left(\L_{N}^{M}|00\rangle \rangle\right)_1 , \\
p_{01}(N,M)&=\left(\L_{N}^{M}|00\rangle \rangle\right)_6 , \\
p_{10}(N,M)&=\left(\L_{N}^{M}|00\rangle \rangle\right)_{11} , \\
p_{11}(N,M)&=\left(\L_{N}^{M}|00\rangle \rangle\right)_{16} , 
%p_{01}(N,M)&=\Big(\L_{N}^{M}|00\rangle \rangle\Big)[6]\\
%p_{10}(N,M)&=\Big(\L_{N}^{M}|00\rangle \rangle\Big)[11]\\
%p_{11}(N,M)&=\Big(\L_{N}^{M}|00\rangle \rangle\Big)[16],
\end{align}
\ese
where the subscripts $1, 6, 11, 16$ denote the index of the obtained column vector elements that correspond to the four basis states $\ket{00},\ket{01},\ket{10},\ket{11} $.

The indirect probabilities for the four basic states as determined by SQPT are illustrated in Fig.~\ref{Fig:Manila_pops_M} with dashed lines. 
These are noticeably different from the directly measured set, attributable to the significant inaccuracies in SQPT arising from SPAM and readout errors.

The curves stemming from \MPPT~utilizing the averaged Liouville matrix $\L_{11}^{M}$ are also presented. 
The corresponding curves for the direct probability measurement and \MPPT~align more closely, providing further evidence that \MPPT~significantly enhances tomographic quality. 
%Additionally, a slight divergence between the linear and iterative method curves is observed, consistent with findings from our simulations.

%%%%%%%%%%%%%% SEC %%%%%%%%%%%%%% SEC %%%%%%%%%%%%%% SEC %%%%%%%%%%%%%% SEC %%%%%%%%%%%%%% 
%%%%%%%%%%%%%% SEC %%%%%%%%%%%%%% SEC %%%%%%%%%%%%%% SEC %%%%%%%%%%%%%% SEC %%%%%%%%%%%%%% 
%%%%%%%%%%%%%% SEC %%%%%%%%%%%%%% SEC %%%%%%%%%%%%%% SEC %%%%%%%%%%%%%% SEC %%%%%%%%%%%%%% 
\section{Conclusions}\label{Conclusions}

In this paper, we introduced a technique called  Multipass Quantum Process Tomography (\MPPT) that greatly improves the precision and accuracy of Standard Quantum Process Tomography by employing gate repetitions (passes). 
This approach serves as a complement to other non-repetitive methods such as SQPT and AAPT, offering a more straightforward and resource-efficient alternative to the more intricate and demanding repetitive methods. 
We explored both iterative and linear versions of \MPPT~and demonstrated their application to the $\sqrt{X}$ and CNOT gates. 
Our simulation models assessed the method's accuracy and precision before their implementation on the real IBM Quantum processor \textsc{ibmq\_manila}. 
The findings indicate that while each of the linear and iterative \MPPT~methods has specific advantages depending on the tomography conditions, both consistently alleviate the effects of SPAM and readout errors, and shot noise. 

We emphasize that, compared to other quantum tomography methods, such as randomized benchmarking, which only deliver the gate fidelity, our method provides the entire quantum process matrix, the Pauli transfer matrix (PTM). 
It contains the complete information for the specific process, which can be used to deduce the contributions of the various error mechanisms.
We achieve this with far better precision and accuracy than the (non-repetitive) Standard Quantum Process Tomography.

%The proposed \MPPT~method can be extended to systems comprising three or more qubits.

\acknowledgements

We acknowledge stimulating discussions with Klaus M{\o}lmer.
This research is supported by the Bulgarian national plan for recovery and resilience, contract BG-RRP-2.004-0008-C01 (SUMMIT: Sofia University Marking Momentum for Innovation and Technological Transfer), project number 3.1.4. and by the European Union’s Horizon Europe research and innovation program under Grant Agreement No. 101046968 (BRISQ).

\appendix
\bwt
\section{Target and Error matrices for the $\sqrt{X}$ and CNOT simulations}\label{A1} 

{In this appendix we present the Target Pauli Transfer matrix $\T$ of $\sqrt{X}$ and CNOT gates and the actual error matrices $\E$ that are utilized for the simulation in Secs. ~\ref{Dem_sim_sx} and ~\ref{Dem_sim_cnot}, respectively. The total process $\R$ is a sum of the corresponding $\T$ and $\E$. }

\be
\T_{\sqrt{X}}=\left[
\begin{array}{cccc}
 1. & 0 & 0 & 0 \\
 0  & 1 & 0 & 0\\
 0 & 0 & 0 & -1 \\
 0 & 0 & 1 & 0 \\
\end{array}
\right] ,
\ee

\be
\E_{\sqrt{X}}=\left[
\begin{array}{cccc}
 0. & 0. & 0. & 0. \\
 8.284 \times10^{-6} & -0.00022872 & 0.00710035 & 0.00693111 \\
 -0.0000204532 & 0.00701596 & -0.0069175 & 0.000281 \\
 0.0000201702 & -0.0069451 & -0.00026103 & -0.00703024 \\
\end{array}
\right] ,
\ee

\be
\T_{CNOT(1,0)}= 
\left[
\begin{array}{cccccccccccccccc}
 1 & 0 & 0 & 0 & 0 & 0 & 0 & 0 & 0 & 0 & 0 & 0 & 0 & 0 & 0 & 0 \\
 0 & 1 & 0 & 0 & 0 & 0 & 0 & 0 & 0 & 0 & 0 & 0 & 0 & 0 & 0 & 0 \\
 0 & 0 & 0 & 0 & 0 & 0 & 0 & 0 & 0 & 0 & 0 & 0 & 0 & 0 & 1 & 0 \\
 0 & 0 & 0 & 0 & 0 & 0 & 0 & 0 & 0 & 0 & 0 & 0 & 0 & 0 & 0 & 1 \\
 0 & 0 & 0 & 0 & 0 & 1 & 0 & 0 & 0 & 0 & 0 & 0 & 0 & 0 & 0 & 0 \\
 0 & 0 & 0 & 0 & 1 & 0 & 0 & 0 & 0 & 0 & 0 & 0 & 0 & 0 & 0 & 0 \\
 0 & 0 & 0 & 0 & 0 & 0 & 0 & 0 & 0 & 0 & 0 & 1 & 0 & 0 & 0 & 0 \\
 0 & 0 & 0 & 0 & 0 & 0 & 0 & 0 & 0 & 0 & -1 & 0 & 0 & 0 & 0 & 0 \\
 0 & 0 & 0 & 0 & 0 & 0 & 0 & 0 & 0 & 1 & 0 & 0 & 0 & 0 & 0 & 0 \\
 0 & 0 & 0 & 0 & 0 & 0 & 0 & 0 & 1 & 0 & 0 & 0 & 0 & 0 & 0 & 0 \\
 0 & 0 & 0 & 0 & 0 & 0 & 0 & -1 & 0 & 0 & 0 & 0 & 0 & 0 & 0 & 0 \\
 0 & 0 & 0 & 0 & 0 & 0 & 1 & 0 & 0 & 0 & 0 & 0 & 0 & 0 & 0 & 0 \\
 0 & 0 & 0 & 0 & 0 & 0 & 0 & 0 & 0 & 0 & 0 & 0 & 1 & 0 & 0 & 0 \\
 0 & 0 & 0 & 0 & 0 & 0 & 0 & 0 & 0 & 0 & 0 & 0 & 0 & 1 & 0 & 0 \\
 0 & 0 & 1 & 0 & 0 & 0 & 0 & 0 & 0 & 0 & 0 & 0 & 0 & 0 & 0 & 0 \\
 0 & 0 & 0 & 1 & 0 & 0 & 0 & 0 & 0 & 0 & 0 & 0 & 0 & 0 & 0 & 0 \\
\end{array}
\right] ,
\ee

\begin{align}
&\E_{CNOT(1,0)} = \notag \\ %10^{-3} \times \\
&\left[ \footnotesize
\begin{array}{cccccccccccccccc}
 0. & 0. & 0. & 0. & 0. & 0. & 0. & 0. & 0. & 0. & 0. & 0. & 0. & 0. & 0. & 0. \\
 -0.01 & -2.22 & -0.36 & 0.61 & 0. & 0. & 0.43 & 0.3 & 0. & 0. & 0.24 & 0.45 & 0. & -0.01 & -9.49 & 9.66 \\
 0.39 & 9.6 & -0.91 & -1.21 & 0.23 & 0.25 & 1.22 & 1.1 & 0. & 0.26 & 1.65 & -0.81 & -0.38 & 0.39 & -4.09 & -9.28 \\
 -0.6 & -9.58 & 1.22 & -0.38 & 0. & -0.25 & -1.1 & 1.21 & -0.2 & -0.22 & 0.81 & 1.65 & -0.61 & -0.59 & 9.36 & -3.56 \\
 0.18 & 0.19 & -0.23 & 0.02 & -0.62 & -9.14 & 0.38 & -0.6 & -0.59 & -9.75 & -0.01 & 0.01 & 8.08 & 1.19 & -0.22 & 0.32 \\
 0.19 & 0.19 & -0.29 & -0.14 & -9.22 & -0.61 & -0.12 & -0.08 & -9.74 & 0.75 & -8.41 & -10.21 & 1.19 & 8.07 & -0.21 & 0.01 \\
 0. & -0.42 & 8.01 & -1.38 & 10.28 & 0.39 & 3.24 & 10.57 & -0.71 & 0.59 & 9.13 & -9.43 & -0.43 & 0.11 & -0.59 & 1.64 \\
 -0.01 & 0.25 & 1.38 & 8.02 & -8.31 & -0.59 & -10.57 & 3.24 & 0.46 & 0.38 & 9.42 & 8.54 & -0.32 & -0.13 & -1.63 & -0.58 \\
 0.13 & -0.12 & 0.02 & 0.2 & 0.72 & 9.99 & 0. & -0.01 & 0.6 & -7.9 & 0.38 & -0.6 & -7.15 & 1.66 & -0.28 & 0.19 \\
 -0.12 & 0.13 & 0.35 & 0.19 & 9.98 & -0.61 & 8.41 & 10.21 & -7.99 & 0.61 & -0.11 & -0.07 & 1.66 & -7.13 & 0.05 & 0.23 \\
 0. & 0.3 & -7.11 & -0.81 & 0.71 & -0.59 & -9.15 & 9.27 & 10.29 & 0.4 & 3.24 & 9.1 & -0.24 & -0.02 & -0.99 & -1.12 \\
 0.01 & -0.44 & 0.81 & -7.12 & -0.46 & -0.38 & -9.26 & -8.56 & -8.33 & -0.6 & -9.1 & 3.23 & -0.44 & 0.04 & 1.13 & -1. \\
 -1.26 & 0. & 0. & 0. & 1.16 & -9.93 & 0.42 & -0.25 & 1.63 & 10.7 & -0.3 & 0.45 & -3.9 & 0.01 & 0. & 0. \\
 -0.01 & -1.27 & -9.5 & 9.65 & -9.93 & 1.14 & 0.05 & 0.07 & 10.7 & 1.61 & 0.14 & 0.12 & 0. & -6.11 & -0.38 & 0.58 \\
 -0.38 & 0.38 & -4.14 & -8.47 & 0.09 & 0.2 & 0.54 & -10.66 & -0.18 & 0.02 & -1.25 & -9.86 & 0.39 & 9.56 & -0.89 & 1.2 \\
 -0.61 & -0.59 & 8.56 & -3.61 & 0.38 & -0.01 & 10.66 & 0.54 & -0.31 & -0.26 & 9.86 & -1.25 & -0.6 & -9.55 & -1.2 & -0.36 \\
\end{array}
\right] \times 10^{-3}.
\end{align}

For the experimental demonstration on \textsc{ibmq\_manila} in Sec. ~\ref{ExpDemManila}, the target CNOT matrix and the measured error matrix are

\be
\T_{CNOT(0,1)}= 
\left[
\begin{array}{cccccccccccccccc}
 1 & 0 & 0 & 0 & 0 & 0 & 0 & 0 & 0 & 0 & 0 & 0 & 0 & 0 & 0 & 0 \\
 0 & 0 & 0 & 0 & 0 & 1 & 0 & 0 & 0 & 0 & 0 & 0 & 0 & 0 & 0 & 0 \\
 0 & 0 & 0 & 0 & 0 & 0 & 1 & 0 & 0 & 0 & 0 & 0 & 0 & 0 & 0 & 0 \\
 0 & 0 & 0 & 1 & 0 & 0 & 0 & 0 & 0 & 0 & 0 & 0 & 0 & 0 & 0 & 0 \\
 0 & 0 & 0 & 0 & 1 & 0 & 0 & 0 & 0 & 0 & 0 & 0 & 0 & 0 & 0 & 0 \\
 0 & 1 & 0 & 0 & 0 & 0 & 0 & 0 & 0 & 0 & 0 & 0 & 0 & 0 & 0 & 0 \\
 0 & 0 & 1 & 0 & 0 & 0 & 0 & 0 & 0 & 0 & 0 & 0 & 0 & 0 & 0 & 0 \\
 0 & 0 & 0 & 0 & 0 & 0 & 0 & 1 & 0 & 0 & 0 & 0 & 0 & 0 & 0 & 0 \\
 0 & 0 & 0 & 0 & 0 & 0 & 0 & 0 & 0 & 0 & 0 & 1 & 0 & 0 & 0 & 0 \\
 0 & 0 & 0 & 0 & 0 & 0 & 0 & 0 & 0 & 0 & 0 & 0 & 0 & 0 & 1 & 0 \\
 0 & 0 & 0 & 0 & 0 & 0 & 0 & 0 & 0 & 0 & 0 & 0 & 0 & -1 & 0 & 0 \\
 0 & 0 & 0 & 0 & 0 & 0 & 0 & 0 & 1 & 0 & 0 & 0 & 0 & 0 & 0 & 0 \\
 0 & 0 & 0 & 0 & 0 & 0 & 0 & 0 & 0 & 0 & 0 & 0 & 0 & 0 & 0 & 1 \\
 0 & 0 & 0 & 0 & 0 & 0 & 0 & 0 & 0 & 0 & -1 & 0 & 0 & 0 & 0 & 0 \\
 0 & 0 & 0 & 0 & 0 & 0 & 0 & 0 & 0 & 1 & 0 & 0 & 0 & 0 & 0 & 0 \\
 0 & 0 & 0 & 0 & 0 & 0 & 0 & 0 & 0 & 0 & 0 & 0 & 1 & 0 & 0 & 0 \\
\end{array}
\right] ,
\ee

\begin{align}
&\E_{Manila} = \notag \\ %10^{-3} \times \\
&\left[ \footnotesize
\begin{array}{cccccccccccccccc}
 0. & 0. & 0. & 0. & 0. & 0. & 0. & 0. & 0. & 0. & 0. & 0. & 0. & 0. & 0. & 0. \\
 1.75 & 2.87 & 2.21 & 1.34 & -2.07 & -11.7 & 0.93 & -5.66 & 4.1 & -8.15 & -2.72 & -1.93 & -6.48 & -25.64 & -0.21 & 0.52 \\
 4.16 & 0.7 & -2.66 & -11.71 & -2.6 & -3.87 & -8.13 & 11.9 & -3.33 & 1.28 & -6.59 & 2.64 & -0.86 & 0.79 & -24.47 & -0.69 \\
 1.83 & 3.37 & -8.85 & -7.86 & -2.44 & -1.92 & 6.38 & 3.45 & -1.06 & 0.99 & -7.82 & 2.32 & 0.02 & -3.42 & -3.03 & -0.53 \\
 5.52 & 2.28 & 5.09 & -1.21 & -12.64 & -1.52 & -4.48 & 1.25 & -9.06 & 3.74 & -4.82 & 2.58 & -26. & -7.35 & -2.81 & 6. \\
 -1.36 & -7.37 & 0.59 & 2. & 2.59 & 3.04 & 0.96 & -5.97 & 0.25 & 0.82 & -4.52 & -4.07 & -1.54 & -0.92 & 4.45 & 2.65 \\
 -4.27 & 2.12 & -11.25 & 12.44 & 4.36 & -3.76 & 5.29 & -10.17 & 0.31 & 4.18 & 3.39 & 6.21 & 0.72 & -4.93 & -0.97 & 3.84 \\
 0.23 & 5.24 & 13.26 & 1.28 & -0.03 & -4.09 & -15.41 & -10.35 & 2.71 & -3.81 & -4.1 & -5.46 & 6.12 & -2.34 & -0.22 & -25.01 \\
 -2.74 & 2.8 & -8.93 & 3.32 & -2.52 & 1.48 & 3.69 & 7.22 & -0.67 & -0.25 & -12.89 & -8.83 & 2.01 & -4.46 & 4.44 & -1.48 \\
 0.69 & -4.1 & -1.76 & 3.79 & 2.66 & -1.41 & 26.68 & 0.76 & -2.45 & 2.27 & 4.88 & 1.9 & -4.42 & -0.14 & -9.72 & 12.48 \\
 3.45 & 0.05 & -4.32 & -3.18 & -7.77 & -25.31 & -0.28 & 2.31 & -10.48 & -3.12 & 1.62 & 3.43 & -1.7 & 9.63 & -2.06 & 0.53 \\
 1.54 & 0.8 & -1.75 & -2.41 & 8.29 & -3.7 & 5.28 & -2.29 & -10.11 & -2.89 & -3.99 & -0.36 & -0.84 & -10.26 & 1.82 & 0.35 \\
 1.67 & -3.23 & -2.6 & -1.7 & -7.9 & 1.61 & -3.34 & 26.74 & -0.91 & 3.59 & 0.41 & 3.76 & 1.24 & 2.55 & -10.44 & -7.56 \\
 -0.2 & -3.79 & -1.99 & -6.29 & -3.32 & 0.06 & -8.02 & 1.44 & 4.42 & 1.33 & 8.5 & -12.88 & -1.5 & -3.61 & 1.72 & 3.04 \\
 0.4 & 2.59 & -6.51 & -1.72 & -2.6 & 7.38 & 2.47 & 0.26 & -4.18 & -9.73 & 1.75 & -1.49 & -10.83 & -1.38 & 0.69 & 5.74 \\
 0.64 & -0.67 & 1.53 & -0.84 & 24.89 & 6.02 & 2.6 & -4.25 & 4.25 & 9.64 & -0.28 & -0.01 & -6.89 & 0.62 & -3.46 & 3.96 \\
\end{array}
\right] \times 10^{-3}.
\end{align}

\begin{align}
\end{align}

\section{Python code for finding the Error matrix}\label{python code}
Here we provide the code used to perform the post processing calculation by solving the equation $\R^N =(\T+\E)^N$, where $\E$ is the unknown error matrix. The remaining are known -- the target matrix $\T$ , total PTM matrix $\R^N$ and passes $N = 2m+1, (m=1,2,...)$ for CNOT and $N = 4m+1, (m=1,2,...)$ for $\sqrt{X}$ gates.
\begin{lstlisting}
import numpy as np
def update_E(T, E, RN, N, alpha):
    """Update E based on the current difference delta."""
    RN_current = np.linalg.matrix_power(T, E, N)
    delta = RN - RN_current
    return E + alpha * delta

def find_E(T, RN, N, alpha=0.01, max_iterations=3000, tolerance=1e-12):
    """Iteratively find E such that (T + E)^N = M."""
    E = np.zeros_like(T)  # Initial guess for E
    for _ in range(max_iterations):
        E = update_E(T, E, RN, N, alpha)
        if np.linalg.norm(np.linalg.matrix_power(T, E, N) - RN) < tolerance:
            break
    return E
\end{lstlisting}
\ewt
%%%%%%%%%%%% BIBLIO %%%%%%%%%%%% BIBLIO %%%%%%%%%%%% BIBLIO %%%%%%%%%%%% BIBLIO %%%%%%%%%%%%


\begin{thebibliography}{999}

\bibitem{NielsenChuang2012}M. A. Nielsen and I. L. Chuang, \emph{Quantum computation
and quantum information}, Cambridge Uni Press (2000).

\bibitem{ChuangNielsen1997}I. L. Chuang and M. A. Nielsen, %Prescription for experimental determination of the dynamics of a quantum black box
, Journal of Modern Optics 44, 2455 (1997),
arXiv:quant-ph/9610001.

\bibitem{Poyatos1997}J.F. Poyatos, J.I. Cirac, and P. Zoller, Phys. Rev. Lett. 78, 390 (1997).
%Complete characterization of a quantum process: The two-bit quantum gate

\bibitem{O'Brien2004}J.L. O'Brien, G.J. Pryde, A. Gilchrist, D.F.V. James,
N.K. Langford, T.C. Ralph, and A.G. White, Phys.
Rev. Lett. 93, 080502 (2004).
%Quantum process tomography of a controlled-NOT gate,

\bibitem{Riebe2006}M. Riebe, K. Kim, P. Schindler, T. Monz, P. O. Schmidt, T. K. Körber, W. Hänsel, H. Häffner, C. F. Roos, and R. Blatt
Phys. Rev. Lett. 97, 220407, (2006)
%Process Tomography of Ion Trap Quantum Gates

\bibitem{Tinkey2021}H. N. Tinkey et al, Quantum Sci. Technol. 6 034013, (2021)
%Quantum process tomography of a Mølmer-Sørensen gate via a global beam

\bibitem{Bialczak2010}Bialczak et al. Nature Phys 6, 409–413 (2010)
% Quantum process tomography of a universal entangling gate implemented with Josephson phase qubits.

\bibitem{Fiurasek2001}J. Fiurasek and Z. Hradil
Phys. Rev. A 63, 020101(R), (2001)
%Maximum-likelihood estimation of quantum processes

\bibitem{Altepeter2003}J. B. Altepeter, D. Branning, E. Jeffrey, T. C. Wei, P. G. Kwiat, R. T. Thew, J. L. O’Brien, M. A. Nielsen, and A. G. White,
Phys. Rev. Lett. 90, 193601 (2003)
%Ancilla-Assisted Quantum Process Tomography

\bibitem{Mohseni2008}M. Mohseni, A.T. Rezakhani, and D.A. Lidar,  Phys. Rev. A 77, 032322 (2008).
%Quantum process tomography: Resource analysis of different strategies,

\bibitem{Roncallo2024}S. Roncallo et al  Quantum Sci. Technol. 9 015010 (2024)
%Pauli transfer matrix direct reconstruction: channel characterization without full process tomography

\bibitem{Rodionov2014} V. Rodionov et al.,
Phys. Rev. B 90, 144504 (2014)
%Compressed sensing quantum process tomography for superconducting quantum gates

\bibitem{Ahmed2023}S. Ahmed, F. Quijandria, and A. F. Kockum,
Phys. Rev. Lett. 130, 150402 (2023)
%Gradient-Descent Quantum Process Tomography by Learning Kraus Operators

\bibitem{Merkel2013}S. T. Merkel et al., Phys. Rev. A 87, 062119, (2013)
%Self-consistent quantum process tomography

\bibitem{Sugiyama2021}T. Sugiyama et al.
Phys. Rev. A 103, 062615, (2021)
%Self-consistent quantum tomography with regularization

\bibitem{Surawy-Stepney2022}Surawy-Stepney et al., Quantum 6, 844 (2022)
%Projected Least-Squares Quantum Process Tomography

\bibitem{Kunjummen2023}J. Kunjummen et al., Phys. Rev. A 107, 042403 – (2023)
%Shadow process tomography of quantum channels

\bibitem{Gulliksen2015}J. Gulliksen, D.B.R. Dasari, K. Mølmer,  EPJ Quantum Technol. 2, 4 (2015).
%Characterization of how dissipation and dephasing errors accumulate in quantum computers

\bibitem{Chow2012a}J.M. Chow et al., Phys. Rev. Lett. 109, 060501, (2012)
%Universal Quantum Gate Set Approaching Fault-Tolerant Thresholds with Superconducting Qubits

\bibitem{Knill2008}E. Knill, D. Leibfried, R. Reichle, J. Reichle, R.B.
Blakestad, J.D. Jost, C. Langer, R. Ozeri, S. Seidelin,
and D.J. Wineland,  Phys. Rev. A 77, 012307 (2008).
%Randomized benchmarking of quantum gates

\bibitem{Proctor2017}T. Proctor, K. Rudinger, K. Young, M. Sarovar, and R. Blume-Kohout
Phys. Rev. Lett. 119, 130502, (2017)
%What Randomized Benchmarking Actually Measures

\bibitem{Magesan2011}E. Magesan, J. M. Gambetta, and J. Emerson,  Phys. Rev. Lett. 106, 180504, (2011)
%Scalable and Robust Randomized Benchmarking of Quantum Processes

\bibitem{Emerson2005}Joseph Emerson et al J. Opt. B: Quantum Semiclass. Opt. 7 S347, (2005 )
%Scalable noise estimation with random unitary operators

\bibitem{Magesan2012}E. Magesan, J. M. Gambetta, and J. Emerson, Phys. Rev. A 85, 042311, (2012)
%Characterizing quantum gates via randomized benchmarking

\bibitem{Corcoles2013} A. D. Córcoles, Jay M. Gambetta, Jerry M. Chow, John A. Smolin, Matthew Ware, Joel Strand, B. L. T. Plourde, and M. Steffen,
Phys. Rev. A 87, 030301(R), (2013)
%Process verification of two-qubit quantum gates by randomized benchmarking

\bibitem{Cross2019}A. W. Cross, L. S. Bishop, S. Sheldon, P. D. Nation, and J. M. Gambetta,
Phys. Rev. A 100, 032328 (2019)
%Validating quantum computers using randomized model circuits

\bibitem{Erhard2019}Erhard, A., Wallman, J.J., Postler, L. et al., Nat Commun 10, 5347 (2019)
%Characterizing large-scale quantum computers via cycle benchmarking

\bibitem{Magesan2012a}E. Magesan et al.,Phys. Rev. Lett. 109, 080505, (2012)
%Efficient Measurement of Quantum Gate Error by Interleaved Randomized Benchmarking

\bibitem{Greenbaum2015}D. Greenbaum, \emph{Introduction to Quantum Gate Set Tomography}, arXiv:1509.02921 [quant-ph], (2015)

\bibitem{Nielsen2021}E. Nielsen et al., \emph{Gate Set Tomography}, 	Quantum 5, 557 (2021)

\bibitem{Chow2012b}J.M. Chow et al.,
 arXiv:1202.5344v1 [quant-ph],(2012) 
%Universal Quantum Gate Set Approaching Fault-Tolerant Thresholds with Superconducting Qubits

\bibitem{Vitanov2020} N. V. Vitanov, New Journal of Physic \textbf{22}, 023015 (2020).
%Relations between single and repeated qubit gates: coherent error amplification for high-fidelity quantum-gate tomography

\bibitem{Stanchev2024} S. G. Stanchev and N. V. Vitanov, Phys. Rev. A 109, 012605 (2024)
%Characterization of high-fidelity Raman qubit gates

\bibitem{Stanchev2023} S. G. Stanchev and N. V. Vitanov,
J. Phys. B: At. Mol. Opt. Phys. 56 014001 (2023).
%Coherent interaction of multistate quantum systems possessing the Wigner-Majorana and Morris-Shore dynamic symmetries with pulse trains

\bibitem{Vitanov1995}
N. V. Vitanov and P. L. Knight, Phys. Rev. A \textbf{52}, 2245 (1995).
%Coherent excitation of a two-state system by a train of short pulses

\bibitem{Vitanov2018} N. V. Vitanov, Phys. Rev. A \textbf{97}, 053409 (2018).
%Relations between the single-pass and double-pass transition probabilities in quantum systems with two and three states 

\bibitem{Baldwin2014}C. H. Baldwin, A. K., and I. H. Deutsch
Phys. Rev. A 90, 012110, (2014)
%Quantum process tomography of unitary and near-unitary maps

\bibitem{Zhou2015}Xiao-Qi Zhou et all,  Optica 2, 510-516 (2015)
%Quantum-enhanced tomography of unitary processes

\bibitem{Sorensen2000}A. Sørensen and K. Mølmer, Phys. Rev. A 62, 022311, (2000)
%Entanglement and quantum computation with ions in thermal motion

\bibitem{Jamiolkowski1972}A. Jamiolkowski, Rep.
Math. Phys. 3, 275 (1972).
%Linear transformations which preserve trace and positive semidefiniteness of operators,

\bibitem{Choi1972} M. D. Choi, Canadian Journal of Mathematics, Volume 24, Issue 3, June 1972, pp. 520 - 529
%Positive Linear Maps on C*-Algebras

\bibitem{Choi1975} M. D. Choi, Linear
Algebra and Its Applications, 10(3), 285–290 (1975).
%Completely positive linear maps on complex matrices.

\bibitem{Lindblad1976}G. Lindblad, Commun.Math. Phys. 48, 119, (1976)
%On the generators of quantum dynamical semigroups

\bibitem{Gorini1976}V. Gorini, A. Kossakowski, E. C. G. SudarshanJ. Math. Phys. 17, 821–825 (1976)
%Completely positive dynamical semigroups of N‐level systems

\bibitem{Kraus1983}K. Kraus, \emph{States, Effects, and Operations Fundamental Notions of Quantum Theory}, Springer (1983)

\bibitem{Stinespring1955}Stinespring, W. Proceedings of the American Mathematical Society, 6(2), 211–216, (1955).
%Positive functions on C*-algebras.

\bibitem{Gambetta2013}J. M. Gambetta, \emph{“Control of Superconducting Qubits.” in Quantum Information Processing}, Lecture Notes of the44th IFF Spring School 2013, edited by D. DiVincenzo(2013) Chap. B4.

\bibitem{Wood2015}C. J. Wood, J. D. Biamonte, and D. G. Cory, 
%Tensor networks and graphical calculus for open quantum systems,"
Quantum Information and Computation 15, 759 (2015), arXiv:1111.6950.

\bibitem{Forest2019}
%Combes et al.
K. Gulshen et al., \emph{Forest Benchmarking: QCVV using PyQuil}. https://github.com/
rigetti/forest-benchmarking  (2019).

\bibitem{Kitaev1997}A Y. Kitaev 1997 Russ. Math. Surv. 52 1191, (1997)
%Quantum computations: algorithms and error correction

\bibitem{Aharonov1998}Aharonov et al. "Quantum Circuits with Mixed States". Proceedings of the Thirtieth Annual ACM Symposium on Theory of Computation (STOC). pp. 20–30 (1998).

\bibitem{Watrous2009}J. Watrous, Theory Comput. 5, 217 (2009).
%Semidefinite programs for completely bounded norms

\bibitem{Watrous2013}J. Watrous, Chic. J. Theoret. Comput. Sci. 8, 1 (2013).
%Simpler semidefinite programs for completely bounded norms

\bibitem{Qiskit}Qiskit contributors, Qiskit: An Open-source Framework for Quantum Computing, (2023)

\bibitem{Qutip2013}J. R. Johansson, P. D. Nation, and F. Nori %: "QuTiP 2: A Python framework for the dynamics of open quantum systems."
, Comp. Phys. Comm. 184, 1234 (2013)

\bibitem{Hashim2023}Hashim, A., Seritan, S., Proctor, T. et al., npj Quantum Inf 9, 109 (2023).
%Benchmarking quantum logic operations relative to thresholds for fault tolerance.

\bibitem{Boulfelfel1994}D. Boulfelfel, et al., IEEE Transactions on Nuclear Science, vol. 41, no. 5, pp. 1746-1754, (1994),
%Three-dimensional restoration of single photon emission computed tomography images.

\bibitem{Song2011}
Caiqin Song, Guoliang Chen,
%An efficient algorithm for solving extended Sylvester-conjugate transpose matrix equations,
%https://doi.org/10.1016/j.ajmsc.2011.03.003
Arab Journal of Mathematical Sciences \textbf{59, 2}, 115-134 (2011)


\bibitem{Noise2024}\url{https://qiskit.org/ecosystem/aer/tutorials/3_building_noise_models.html}, Accessed (2024-01-27)

%Building Noise Models - Qiskit Aer 0.13.1
\bibitem{McKay2017}David C. McKay, et al. Phys. Rev. A 96, 022330, (2017)
%Efficient  Z gates for quantum computing

\bibitem{MPT2024}\url{https://qiskit.org/ecosystem/experiments/stubs/qiskit_experiments.library.tomography.MitigatedProcessTomography.html}, Accessed (2024-01-27)
%MitigatedProcessTomography — Qiskit Experiments 0.5 0.5.4 documentation

\bibitem{Gruber1998}M. H. Gruber ,Improving efficiency by shrinkage: the James-Stein and ridge regression estimators, (1998),ISBN 978-0-8247-0156-7

\bibitem{IBM} IBM Quantum (2023). \textit{ibmq\_manila (Falcon r5.11L)}  \url{https://quantum.ibm.com/}


\end{thebibliography}
\end{document}